\documentclass[10pt,a4paper,final]{iopart}   

\expandafter\let\csname equation*\endcsname\relax
\expandafter\let\csname endequation*\endcsname\relax
\usepackage{iopams}  
\usepackage{graphicx}
\usepackage{float}
\usepackage{url}
\usepackage{physics}
\usepackage{xcolor} 
\usepackage[breaklinks=true,colorlinks=true,linkcolor=blue,urlcolor=blue,citecolor=blue]{hyperref}
\usepackage{subfig}
\usepackage{csquotes}

\begin{document}


\title[Ab-initio dielectric response function of diamond]{Ab-initio dielectric response function of diamond and other relevant high pressure phases of carbon} 

\author{Kushal Ramakrishna$^{1,2}$}
\address{$^1$Helmholtz-Zentrum Dresden-Rossendorf, Bautzner Landstra\ss e 400, 01328 Dresden, Germany}
\address{$^2$Technische Universit\"at Dresden, 01062 Dresden, Germany}
\ead{k.ramakrishna@hzdr.de}

\author{Jan Vorberger$^{1}$} 
\address{Helmholtz-Zentrum Dresden-Rossendorf, Bautzner Landstra\ss e 400, 01328 Dresden, Germany}
\ead{j.vorberger@hzdr.de} 


\begin{abstract}
The electronic structure and dielectric properties of the diamond, body centered cubic diamond (bc8), and hexagonal diamond (lonsdaleite) phases of carbon are computed using density functional theory and many-body perturbation theory with the emphasis on the excitonic picture of the solid phases relevant in the regimes of high-pressure physics and warm dense matter. We also discuss the capabilities of reproducing the inelastic x-ray scattering spectra in comparison with the existing models in light of recent x-ray scattering experiments on carbon and carbon bearing materials in the Megabar range.

\end{abstract}

\vspace{2pc} 
\noindent{\it Keywords}: high pressure effects, dielectric functions, warm dense matter
\maketitle



\section{Introduction} 
\par 
As one of the most abundant elements in the universe, carbon plays a unique role with its ability to form hybrid bonds leading to a wide range of allotropes and compounds. Carbon, subjected to extreme pressure and temperature conditions inside the earth and other planets experiences drastic changes in its material properties. Diamond rain is theorized to occur deep inside the layers of Neptune and Uranus due to the demixing of the carbon and hydrogen atoms of methane under conditions of high pressure and temperatures~\cite{Ross1981}. Such conditions can be recreated in the laboratory using compression methods like laser driven shock waves to such an effect that nano-diamonds in a high pressure fluid of carbo-hydrates can indeed be observed~\cite{Kraus2017,doi:10.1063/1.5017908}. Meanwhile computer simulations need to provide significant information prior to the design of such experiments and are essential for the evaluation of the measurements~\cite{doi:10.1063/1.5017908,Hartley2018}.
\par
For inertial confinement fusion applications (ICF), the design of targets is crucial for the feasibility. Diamond is a promising candidate for the ablator material due to it's high atomic density, yield strength and chemical inertness. The understanding of the high pressure solid phase bc8 is important in ICF with a diamond ablator as the shock sequences are needed to be performed avoiding the bc8 phase. Plastic ablator materials are also of interest for ICF and the phase separation of carbon and hydrogen during the implosion could lead to hydrodynamic instabilities with reduced performance in the implosion~\cite{Knudson1822, 0029-5515-49-11-112001, Hurricane2014}.  
\par
Under meteorite impacts, a variety of new high pressure phases of carbon have been reported. Theoretically, these phases are found to be metastable under ambient conditions ~\cite{FERROIR2010150}. Lonsdaleite or hexagonal diamond is one such phase discovered for the first time in 1967 in the remains of a meteorite and later synthesized under laboratory conditions~\cite{Lons-1967}. The formation is reported in a wide variety of experiments with the veracity of the recovery under ambient conditions ambiguous~\cite{doi:10.1063/1.1841236, PhysRevB.46.6031, 0953-8984-16-14-011,  Nemeth2014, Kraus2016, Turneaureeaao3561}.
\par
The allotropes of carbon feature a wide range of properties such as contrasting hardness from graphite to diamond~\cite{PhysRevB.83.193410}. The potential application of graphene is already widely known and diamond is one of the most studied wide-band semiconducting material~\cite{Geim1530}. Theoretically, lonsdaleite and bc8 are harder than diamond and the knowledge of electronic and optical properties is important for characterizing their potential applications~\cite{doi:10.1063/1.5025856}. The stable bc8 and lonsdaleite phases of silicon and germanium have already been synthesized using diamond anvil cells and laser ablation~\cite{Wentorf338, HU1984263, OLIJNYK1984137, PhysRevB.34.4679, 0022-3727-49-28-285304}. The bc8 phase of silicon is a narrow direct band gap semiconductor at room temperature and the lonsdaleite phase is the stabler form for synthesizing nanowires under certain dimensions~\cite{PhysRevLett.118.146601, PhysRevLett.95.115502}.
\par
The description of the electronic and ionic properties of carbon, of the equation of state (EOS), and the corresponding phase boundaries requires state of the art ab initio methods such as density functional theory (DFT) with molecular dynamics (MD) or path integral Monte Carlo. A combination of DFT and many-body quantum statistics allows to include higher order correlations and describe new physics such that the accuracy and predictive power of theoretical methods is enhanced. We are particularly interested in the dynamic structure factor (DSF) as it is an important quantity to determine properties of high pressure solids and fluids and warm dense matter states. The DSF can be accessed, e.g., using energetic x-ray free-electron (XFEL) radiation or electron beams. Enhanced DSF models include in particular electron-hole interactions in semiconductors and insulators. This provides improved predictions for dielectric functions and conductivities, especially when it is combined with higher rungs of xc-functionals for a better description of the band gaps. From these, better EOS may be derived based on the DSF or dielectric function.  As a practical application, the determination of the temperature based on x-ray Thompson scattering (XRTS) data will be improved \cite{RevModPhys.81.1625}.

\section{Theory of inelastic x-ray scattering}

\par
The density fluctuations off which x-rays scatter can be described by the DSF $S(q,\omega)$ using the density response function $\chi(q,\omega)$ with the inverse of the dielectric function $\varepsilon^{-1}(q,\omega)=1+V(q)\chi(q,\omega)$ where $V(q)$ is the Coulomb potential via the fluctuation-dissipation theorem 

\begin{equation}
\centering
S(q,\omega) = \frac{ \hbar  }{ \pi  n_{e} } \frac{1}{ 1 - e^{ \hbar \omega / k_{B} T_{e}  } } \Im[ \chi(q,\omega)].
\end{equation} 

Here, $n_{e}$ is the number of free electrons and $q=||\vec{q}||$ the magnitude of the scattering vector. This also obeys the detailed balance relation, $ S(q,\omega) = S(-q,-\omega)  e^{ - \beta \hbar \omega } $ and the asymmetry with respect to \emph{q} and $\omega$ is used to infer the electron temperature off the plasmons in experiments~\cite{RevModPhys.81.1625, PhysRevLett.115.115001}. The resonance frequency $\omega$ used in the detailed balance equation is up (down)-shifted $\omega \rightarrow |\omega_{\boldsymbol{X}} \pm \omega|$ based on the energy gain or loss of the x-rays where $\omega_{\boldsymbol{X}}$ is the frequency of the x-ray source. The extended Mermin ansatz (MA) by Fortmann \etal provides the state of the art description of the free electron feature of the scattering signal in warm dense matter (WDM)~\cite{PhysRevB.1.2362, PhysRevE.81.026405}. It allows to consider electron-electron correlations and also the influence of electron-ion collisions by the inclusion of dynamical collision frequency and local field corrections (LFC). This approach also allows the application to XRTS for degenerate plasmas at temperatures far below the Fermi temperature. The electron-electron response function under MA is given by

\begin{eqnarray}
\lefteqn{\chi^{M}(q,\omega) =}&&\label{MA}\\
&& \left(1 - \frac{i\omega}{\nu(\omega)}\right) \frac{ \chi^{LFC}\left(q,\omega + i \nu(\omega)\right)\chi^{LFC}(q,0)  }{ \chi^{LFC}\left(q,\omega + i \nu(\omega)\right) - \frac{i\omega}{\nu(\omega)} \chi^{LFC}(q,0) },\nonumber
\end{eqnarray} 

where $\chi^{LFC}(q,\omega)$ is the single component response function with the inclusion of dynamical local field correction $G(q,\omega)$ and $\nu(\omega)$ is the collision frequency. $\chi^{LFC}(q,\omega)$ is given by

\begin{equation} 
\chi^{LFC}(q,\omega) = \frac{ \chi^{0}(q,\omega) }{1 - V(q)\left[1-G(q,\omega)\right]\chi^{0}(q,\omega) ) },
\end{equation}

and setting $G(q,\omega)=0$ leads to the level of RPA~\cite{ROPKE1999365, Selchow_2002}. 
The free density response functions $\chi^0$ has usually been calculated using free electron states (and the density of free electrons) for applications in plasma and warm dense matter physics.
The complex collision frequency $\nu(\omega)$ needed for the Mermin approach of Eq. (\ref{MA}) may in WDM be taken in Born  approximation~\cite{PhysRevE.62.5648}. Recently, it was suggested to extract it directly from DFT using $\varepsilon^{M}(q\rightarrow0,\omega)=\varepsilon^{DFT}(q\rightarrow0,\omega)$ and the Kubo-Greenwood (KG) formula~\cite{PhysRevLett.118.225001, 1367-2630-14-5-055020}. We use this approach as well in order to extend KG-DFT results via MA to finite wavenumbers.
\par
However, using single particle Kohn-Sham states $\psi({\bf r})$ as computed via DFT in the free density response function~\cite{marques:2004}
\begin{equation}
\chi^0_{KS}({\bf r},{\bf r}',\omega)=\lim_{\eta\to 0^+}\sum_{jk}\left(f_k-f_j\right)
\frac{\psi_j({\bf r})\psi_j^*({\bf r}')\psi_k({\bf r}')\psi_k^*({\bf r}')}
{\omega-(\epsilon_j-\epsilon_k)+i\eta},
\label{chi0}
\end{equation}
allows to obtain the density response function within linear response based on time-dependent DFT (TDDFT) 
\begin{equation}
\chi^{TDDFT}(q,\omega) = \frac{ \chi^{0}_{KS}(q,\omega) }{1 - \left[V(q)-f_{xc}(q,\omega)\right]\chi^{0}_{KS}(q,\omega) ) }.
\end{equation} 
Here, the kernel $f_{xc}$ describes higher order exchange and correlation and is, similarly to the local field corrections, subject of active research~\cite{PhysRevB.69.155112}. We can compare these TDDFT calculations that can provide \emph{q}-dependent response functions to the MA based on Kubo-Greenwood-DFT collision frequencies discussed in section \ref{comp_methods}. 
Recently, TDDFT was first used for modeling experimental XRTS spectra of metals. The DSF for warm dense beryllium without the Chihara formalism was computed with the aid of real-time TDDFT and Mermin's formalism for finite-temperature DFT~\cite{PhysRevLett.116.115004}. Mo \etal have used TDDFT to infer the electronic temperature based on the XRTS spectrum of the experiments on aluminium by Sperling \etal~\cite{PhysRevLett.120.205002, PhysRevLett.115.115001}.  
 
Further, more systematic approximations for the structure factor may be derived on the basis of the Bethe-Salpeter equation (BSE)~\cite{newbook}
\begin{eqnarray}
\lefteqn{L(12,1'2')=L^0(12,1'2')+\int_{\cal C}d3d3'd4d4'\,L^0(13,1'3')}&&\\
&&\times\left[
\delta(3-3')\delta(4-4')V(3'-4')+\frac{\delta\bar{\Sigma}(3',3)}{\delta g(4,4')}
\right]
L(42,4'2')\,.\nonumber
\end{eqnarray}
Here, $1=\{{\bf r}_1,t_1,\sigma_1\}$ is a full set of observables and $L(12,1'2')$ is the correlation function of density fluctuations which simplifies to the density response function $\chi(1,2)=L(12,1^+2^+)$ upon setting the times $t_1'=t_1^+$ and $t_2'=t_2^+$. $\bar{\Sigma}(3',3)$ is the screened self-energy and $g(4,4')$ is the single-particle Green's function. $L^0$ denotes the free correlation function akin to the function given in Eq. (\ref{chi0}) once the particle-hole channel is selected. Within this paper, the screened self-energy is taken in GW approximation and only the contribution of the screened potential is considered in the functional derivative. Further, only static screening is taken into account to simplify the time structure of the BSE.    

\section{Computational methods}
\label{comp_methods}

\par
The density functional theory (DFT) calculations along with the random phase approximation (RPA), the TDDFT and BSE computations were primarily performed using a full-potential linearized augmented-plane wave code implemented in elk~\cite{elk}. While DFT has been successful in describing  ground state properties, the excited state properties are not well described by Kohn-Sham eigenvalues especially in the realm of linear response functions and band gaps for insulators and semiconductors ~\cite{PhysRev.140.A1133}. Within the framework of DFT, higher order correlations and exchange contributions leading to e.g., more accurate band gaps, can be accounted for by using xc functionals in various higher rungs of Jacob's ladder ~\cite{perdew:2010}. Beyond pure DFT, the GW approximation is available. The GW approximation leads to a frequency dependency on the GW (Montroll-Ward) self-energy terms thus including medium effects on the electronic states providing improved band gaps~\cite{perdew:2010, 0034-4885-61-3-002}. For GW or single-shot GW$_{0}$ corrections to the band gap, we use the inbuilt GW routines in VASP~\cite{PhysRevB.47.558, PhysRevB.59.1758, KRESSE199615, PhysRevB.54.11169} and yambo~\cite{MARINI20091392} with the Kohn-Sham wavefunctions generated using Quantum ESPRESSO~\cite{QE-2009,QE-2017}.  
\par
The Kubo-Greenwood (KG) formula used ubiquitously under WDM conditions for linear response calculations of the dielectric function and conductivity in the optical limit ($q\to 0$) generally provides results of unknown accuracy only, especially for (partially) bound state systems due to the combination of single state DFT wavefunctions along with the lack of  many-body physics in the formula itself. The KG expression for the frequency dependent conductivity tensor depends on the transition matrix elements $\bra{i} \nabla \ket{j}$ and is given by

\begin{equation}
\centering
\sigma(\omega) = i \frac{2 \hbar e^{2} }{ m^{2}_{e} V} \sum_{i,j} \frac{ \bra{i} \nabla \ket{j} \bra{j} \nabla \ket{i}   }{ E_{i} - E_{j} - \hbar \omega +  i\delta/2 } \frac{ F_{i} - F_{j} }{ E_{i} - E_{j} },
\end{equation}

where $F_{i}$ are the Fermi-Dirac occupations and $E_{i}$ the corresponding single-particle eigenvalues calculated using DFT~\cite{doi:10.1143/JPSJ.12.570, 0370-1328-71-4-306, CALDERIN2017118}. The dielectric function is then calculated using 

\begin{equation}
\centering
\varepsilon(\omega) = 1 + \frac{i}{ \epsilon_{0} \omega } \sigma(\omega).
\end{equation}

\par 
To calculate response functions beyond the level of the KG formula, we employ various many-body methods that take wavefunctions from DFT as input. The TDDFT calculations under adiabatic local density approximation (ALDA) for the xc kernel are performed using Bootstrap, a long range xc kernel implemented in elk, which is reasonably good for reproducing excitonic effects and computationally fast for ab-initio calculation of absorption spectra~\cite{marques:2004,PhysRevB.69.155112,PhysRevLett.107.186401}. RPA uses electron states from DFT, but lacks the xc kernel as used in TDDFT. RPA provides the next best approximation to the Hartree-Fock approximation, representing a change in electron's self-energy due to dynamical  screening~\cite{PhysRev.85.338, PhysRev.92.609, citeulike:12942339}. The solution of the BSE provides the most systematic description of electron-hole correlations including excitons within the gap~\cite{PhysRev.84.1232}. The solution encompasses a two-step process where the quasiparticle electron states and wavefunctions calculated under the GW approximation are used to solve the BSE using a four-point polarization propagator in a Dyson-like equation~\cite{Interactingelectrons}. In contrast, TDDFT uses two-point propagators and is much easier to solve~\cite{PhysRevB.92.045209}. The interaction between the electron-hole pair is approximated by a Coulomb kernel along with a screened interaction term. BSE needs more \emph{k}-points and empty bands than GW to obtain convergence and scales as $\Or(n^{5})$, it is therefore the most computationally demanding method of all used here~\cite{botti_firstprinciples_book}. Diagonalization scaling is given by $(N_{c} \times N_{v} \times N_{k})^{3}$ where $N_{c}$, $N_{v}$ and $N_{k}$ stands for the number of conduction bands, valence bands, and \emph{k}-points respectively~\cite{0953-8984-26-36-363202,B903676H,PhysRevB.95.155121}. The BSE calculations are performed using the method of diagonalization in elk. The code {\em exciting} was used for the finite wavenumber BSE calculations~\cite{0953-8984-26-36-363202,B903676H}. 
 
\section{\label{sec:level3}Electronic structure and optical response of different carbon phases}

\subsection{Diamond} 

\par
\begin{figure}[th] 
\centering 
\includegraphics[width=0.6\columnwidth]{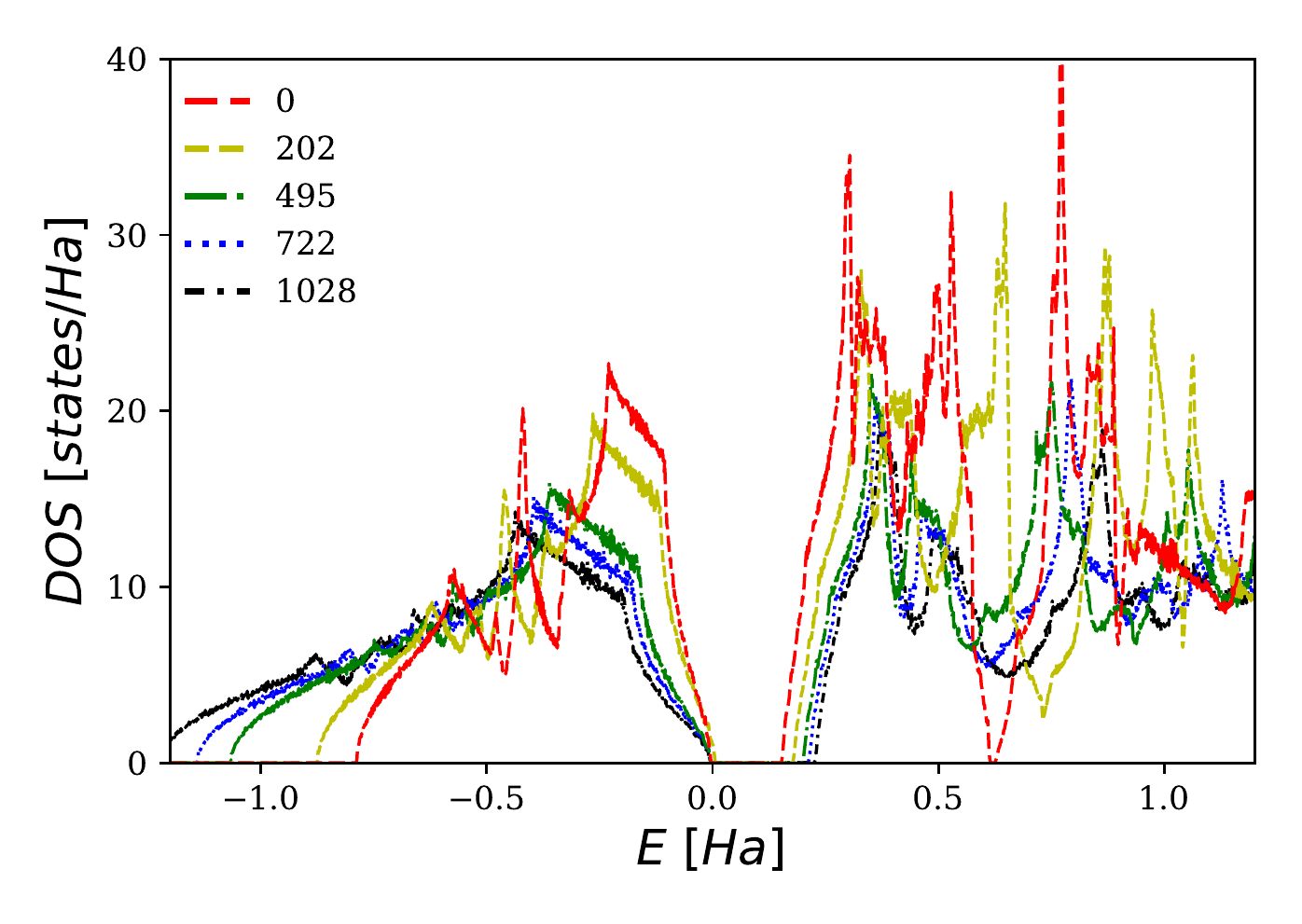}
\caption{ \small{The density of states for diamond as function of pressure obtained from DFT using the all-electron full potential elk code. The pressures are in GPa and the valence band maximum is adjusted to zero.}}
\label{Diamond_fcc_dos} 
\end{figure}

 
Diamond has a face-centered cubic structure with the spacegroup Fd$\bar{3}$m consisting of 2 atoms per primitive unit cell. The lattice parameters are $a=b=c$, $\alpha=\beta=\gamma=90^{0}$. The DFT calculations were performed using elk on a $20\times20\times20$ \emph{k}-point mesh and 16 bands using a PBE-GGA functional with Broyden mixing~\cite{PhysRevB.33.8822, PhysRevLett.77.3865}. The choice of the xc functionals under extreme pressures is important as the system's energy depends on the spatially varying electronic charge densities and the PBE functional has been well tested up to 1000 TPa for the allotropes of carbon~\cite{PhysRevLett.108.045704}. The total energy obtained from DFT for various lattice parameters are fit to the Vinet equation to obtain the equilibrium volume~\cite{0953-8984-1-11-002, PhysRevB.35.1945}, see \ref{app_eos}. The obtained lattice constant a$_{0}$=3.569 \AA \hspace{1 pt} agrees nicely with the experimental measurement 3.567 \AA~\cite{Madelung}. The electronic density of states at various pressures are shown in Fig. \ref{Diamond_fcc_dos}. There is pressure broadening of the valence band, but the most striking property of the diamond electronic structure is the opening of the band gap with increasing pressure. 
 
\par  
It is well known that standard DFT using the LDA or GGA underestimates the band gap. To correct the DFT band gap, we use the GW approximation implemented in VASP with 64 bands on a $16\times16\times16$ \emph{k}-point mesh centered around the Gamma point using the PBE-GGA functional with the hard carbon PAW pseudopotential and the energy cutoff set to 10 Ha. The GW$_{0}$ calculations are performed using the RPA results with the inclusion of local-field effects. 

\begin{figure}[t]
\centering
\includegraphics[width=0.6\columnwidth]{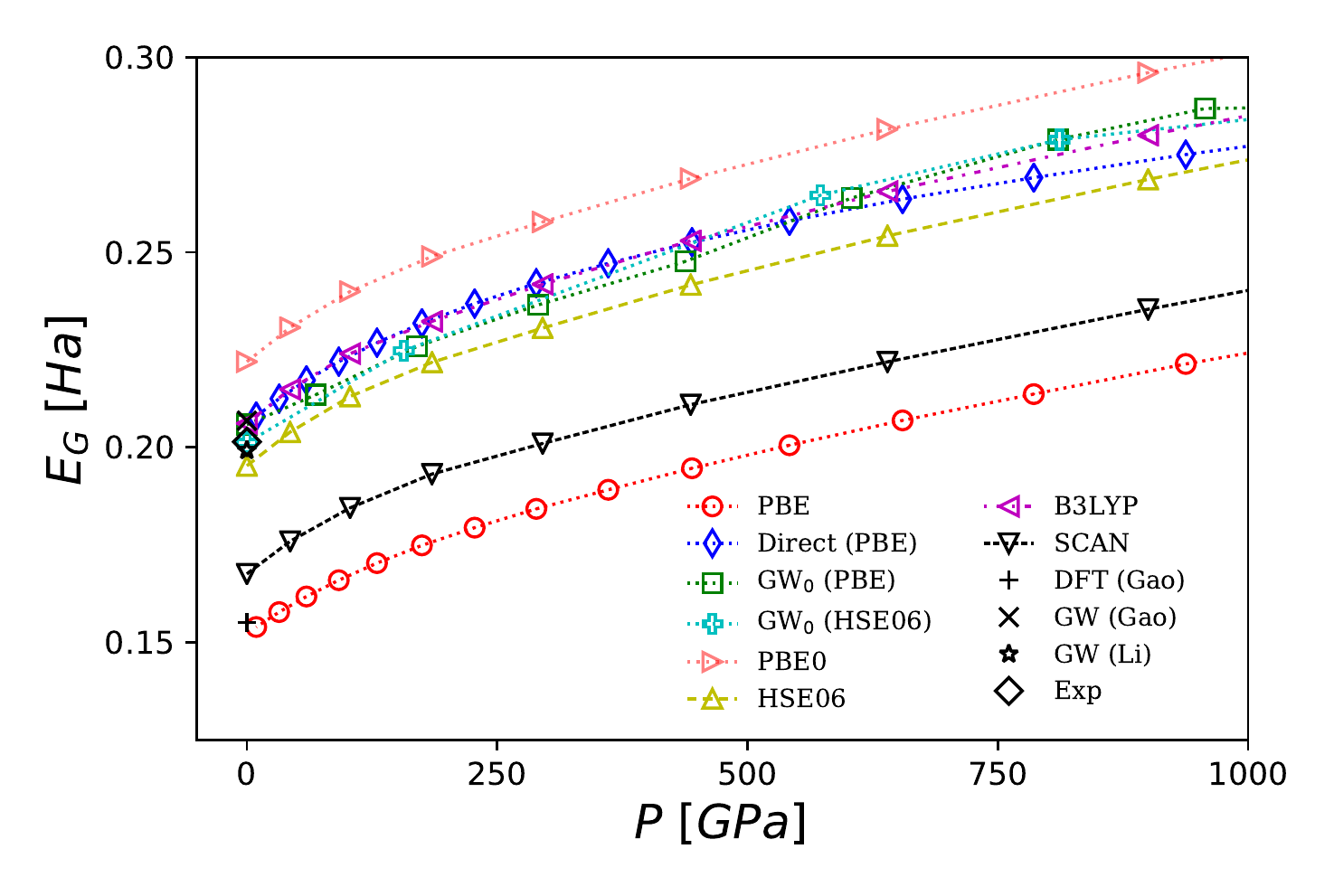} 
\caption{ \small{The band gap of diamond calculated using various xc-functionals as function of pressure in DFT and within the GW approximation. We also show the direct band gap calculated using PBE. Gao~\cite{PSSB:PSSB201451197}; Li~\cite{1367-2630-14-2-023006}; Experiment~\cite{Cardona}. } } 	 
\label{Diamond_fcc_bandgap} 
\end{figure}

Both the direct and the indirect band gaps widen almost linearly with increasing pressures, see Fig. \ref{Diamond_fcc_bandgap}, corroborating the experimental evidence of the band gap opening in diamond under hydrostatic compression by Gamboa \etal~\cite{PhysRevB.35.5856, osti_1241296}. 
As expected, the PBE band gap is smallest and improved approximations, either by advanced functionals or due to many body theory, increase the band gap. Our result for the band gap at zero pressure within the GW$_{0}$ approximation is in the range obtained by other simulations~\cite{PhysRevB.83.193410, PSSB:PSSB201451197, 1367-2630-14-2-023006} and shows good agreement with experiment~\cite{Cardona}. Moving beyond LDA/GGA functionals, especially hybrid functionals for semiconductors and insulators for elements of group-IV have been shown to provide reliable bandgaps~\cite{PhysRevB.78.121201, PhysRevB.80.115205}. Across the pressure ranges considered, HSE06 seems consistently best among the functionals tested for DFT and the GW band gap corrections~\cite{doi:10.1063/1.1564060, PhysRevB.78.121201}. Interestingly, the GW$_0$ results for the band gap do not seem to depend on the choice of xc-functional. While the zero-point energy renormalization is important for finite-temperature electronic effects for carbon materials due to electron-phonon contributions, it can be neglected for the high pressures considered in this work~\cite{PhysRevB.94.075125, PhysRevB.89.214304, 0295-5075-98-6-66007, PhysRevLett.107.255501, PhysRevLett.101.106405}.

\par
\begin{figure}[t]
\centering
\includegraphics[width=0.6\columnwidth]{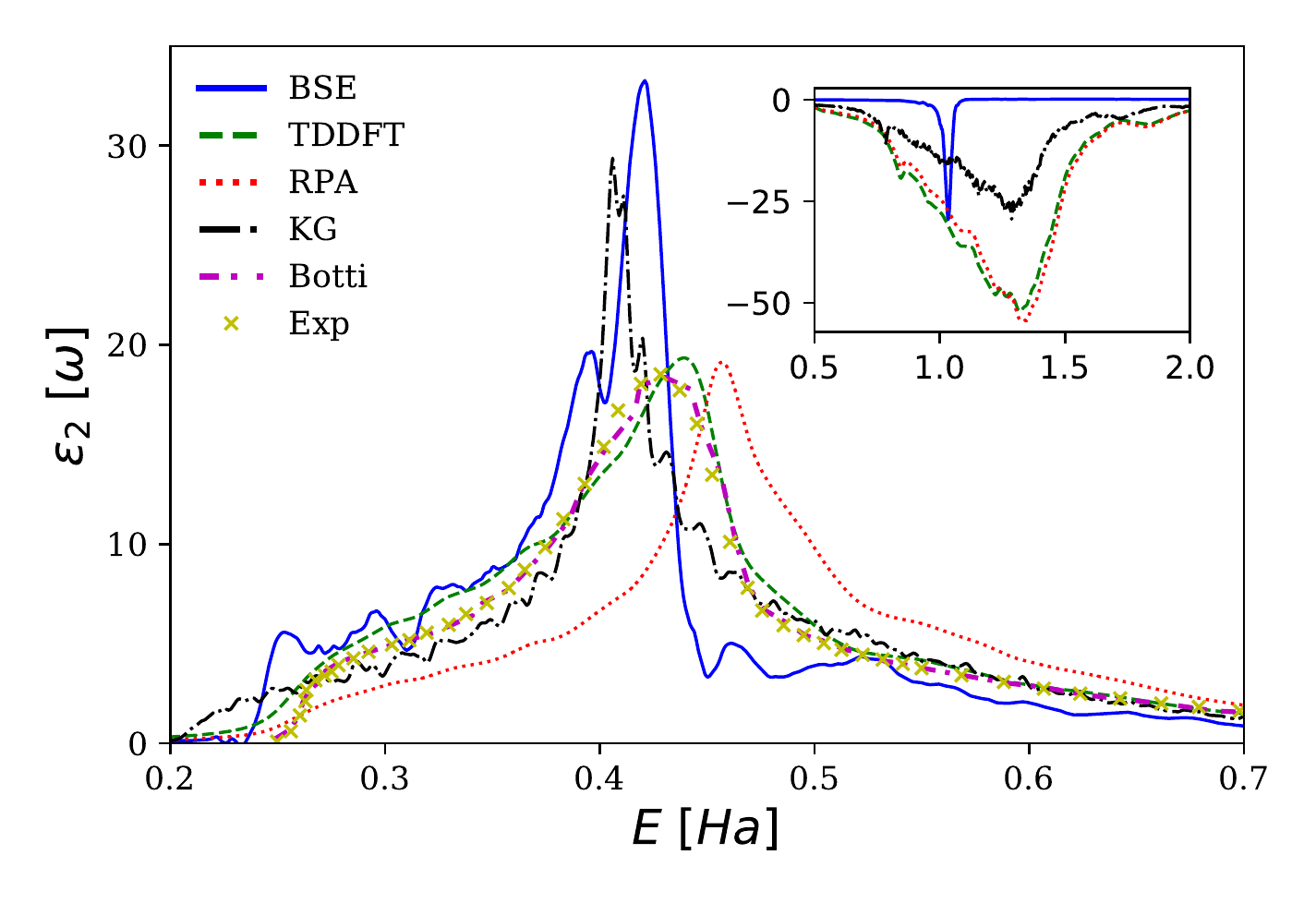} 
\caption{ \small{The imaginary part of the dielectric function of diamond in the optical limit at the equilibrium volume ($\rho=3.509$ g/$cm^{3}$) using various approaches. The inset plot shows the imaginary part of the inverse of the dielectric function including the plasmon peaks. In the inset, the curves obtained using TDDFT/RPA and KG are scaled twentyfold and tenfold, respectively, for the purpose of visualization. The Botti result is taken from Ref.~\cite{PhysRevB.69.155112}; the experimental data stems from Ref.~\cite{Palik:396087}. } }
\label{Diamond_fcc_eps_bse_tddft_rpa}
\end{figure}

 
The response function for diamond evaluated using various approaches is shown in Fig. \ref{Diamond_fcc_eps_bse_tddft_rpa}. Due to the large band gap, excitonic effects are important and therefore BSE is among the methods used to calculate the absorption spectra. For the BSE calculations, we use a $12\times 12 \times12$ \emph{k}-point mesh with 8 empty states. The optical band gap can be deduced from the first peak of the absorption curve (imaginary part of the dielectric function). In some cases there can be a dark state with small oscillator intensities. In general, it is apparent that the onset of absorption happens softer and earlier for KG, RPA, and our TDDFT result than for the BSE model or for the experimental result. The maximum in absorption ranges from $0.407$ Ha for the KG model to $0.465$ Ha in RPA. The experimental maximum is located at $0.427$ Ha and is best reproduced by TDDFT closely followed by BSE.
Though the BSE spectra matches the location of the absorption maxima reasonably well, the oscillator intensity is quite large compared to the experimental result. Such an overestimation is also reported in other calculations~\cite{PhysRevMaterials.1.054603, PhysRevLett.107.186401, PhysRevB.57.R9385}. The overall match to the experimental result seems best for the TDDFT method. The deviation of RPA is understandable as it misses the long range $f_{xc}$ kernel and therefore cannot capture  bound excitons. Thus, the effective result of taking into account the electron-hole interaction in the xc-kernel is a shift of oscillator intensity to lower energies, this effect is prominent for insulators and less for narrow-gap semiconductors~\cite{PhysRevB.57.R9385}. 

\par
\begin{figure}[t]
\centering  
\includegraphics[width=0.6\columnwidth]
{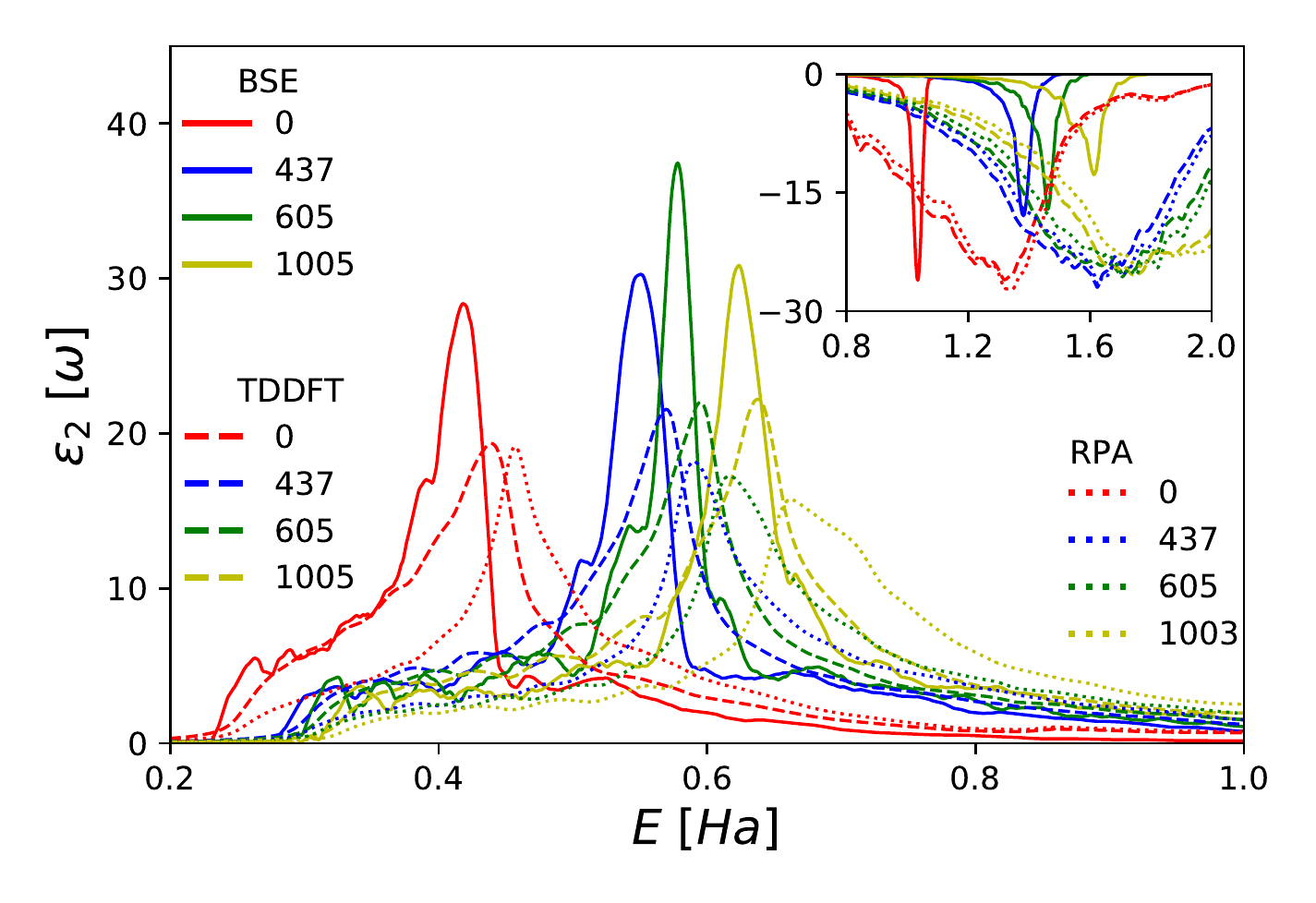}
\caption{ \small{The imaginary part of the dielectric function of diamond in the optical limit for various pressures using BSE (solid), TDDFT (dashed) and RPA (dotted). The inset plot shows the imaginary part of the inverse of the dielectric function including the plasmon peaks. In the inset, the curves obtained using TDDFT and RPA are scaled tenfold for the purpose of visualization. All the pressures indicated are in GPa. The densities for these pressures are $3.51$ g/cc, $5.50$ g/cc, $6.00$ g/cc and $7.00$ g/cc respectively.} } 
\label{Diamond_fcc_bse2}
\end{figure}

At higher pressures, the lower edge of the imaginary part of the dielectric function shifts to higher energies due to the opening of the band gap as shown in Fig. \ref{Diamond_fcc_bse2}. The maximum of the imaginary part of the dielectric function shifts to higher energies even faster than the edge. This is due to the number of occupied and unoccupied states on both sides of the band gap, respectively, declining with an increase in pressure. One observes a similar behaviour of different theories for higher pressures as for the zero pressure case of Fig. \ref{Diamond_fcc_eps_bse_tddft_rpa}. RPA shows the absorption peak at the highest energies, BSE shows the highest absorption strength. TDDFT produces similar absorption spectra as BSE with lower oscillator strengths near the right exciton energies in Fig. \ref{Diamond_fcc_bse2}.

\begin{figure}[t]
\centering 
\includegraphics[width=0.6\columnwidth]{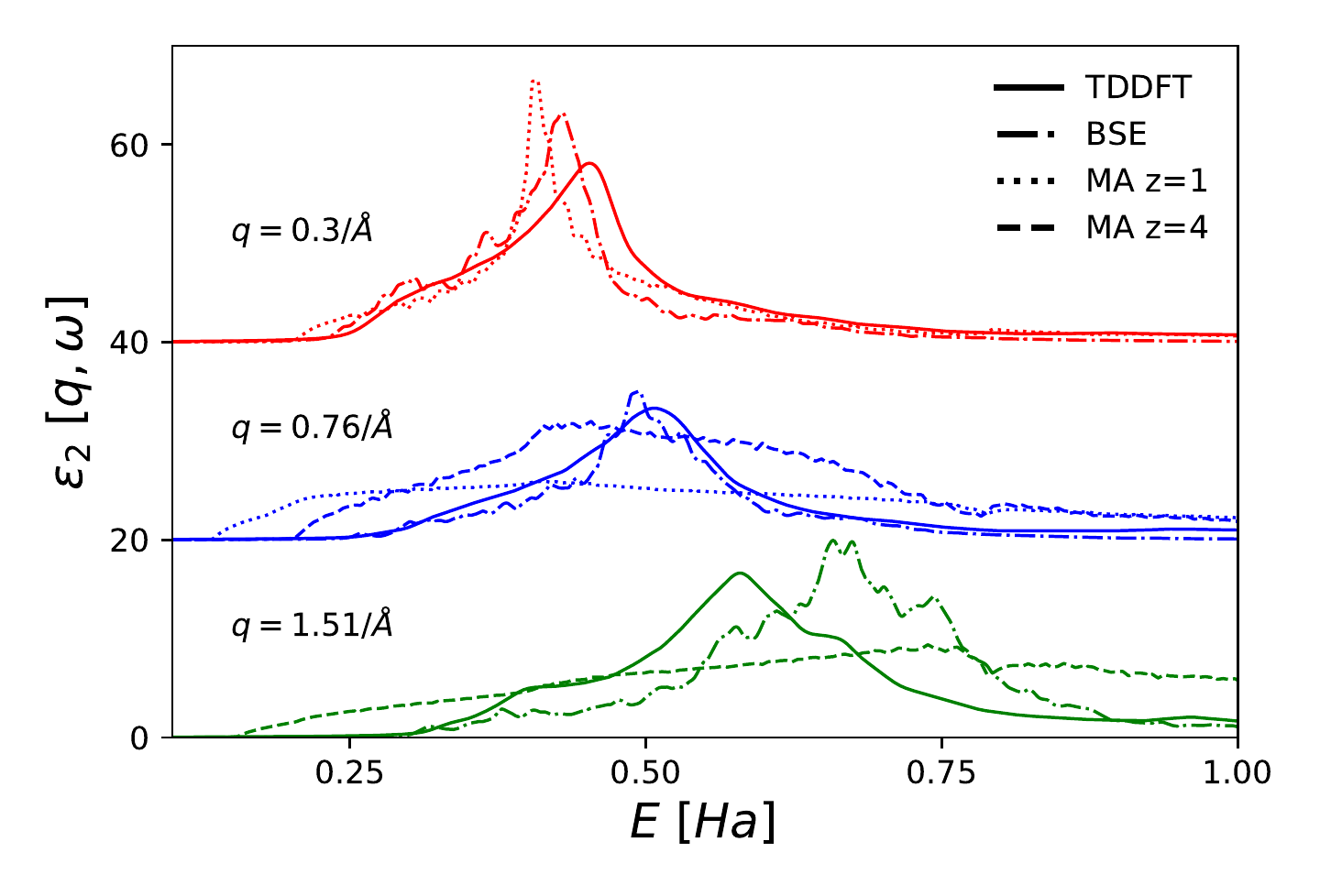} 
\caption{ \small{The imaginary part of the dielectric response function for diamond is calculated using TDDFT, BSE and MA at finite $\vec{q}$ for $\rho=3.509$ g/$cm^{3}$. The charge state $Z=1$ is considered for $q=0.3/ \AA$ and $0.76/ \AA$ and $Z=4$ is considered for $q=0.76/ \AA$ and $1.51/ \AA$.    } }     
\label{Diamond_fcc_eps_bse_tddft_rpa_x}  
\end{figure}

In addition, it is of interest to calculate the dielectric function for other wavenumbers than just in the optical limit $q \rightarrow 0$. Figure \ref{Diamond_fcc_eps_bse_tddft_rpa_x} shows the result of three different theories for finite wavenumbers. 
To choose the complementary \emph{k} and \emph{q} points we use the relation $ n_{k_{i}} \times q_{i} = N$ where $n_{k_{i}}$ is the number of \emph{k}-points along $\vec{x}$, $q_{i}$ is the number of \emph{q}-points along $\vec{x}$, N is an integer which is also a factor of $n_{k_{i}}$ explained by the Monkhorst-Pack \emph{k}-sampling~\cite{PhysRevB.13.5188}. The calculations are performed using an uniform $20 \times 20 \times 20$ \emph{k}-point mesh with 8 empty bands. The TDDFT results were obtained using elk, the BSE results stem from {\em exciting}.  

In general, an increase in momentum leads to a reduction of the peak height and a shift of the peak to higher energies in the imaginary part of the dielectric function. It can be observed that the reduction in peak height as predicted by BSE is much more drastic in the considered q-range as the change in the TDDFT result. Further, the BSE peaks change their location more than the TDDFT peaks. 

TDDFT and BSE derive the wavenumber dependence via differences between \emph{k}-points within the approach. In contrast, a wavenumber dependency of the dielectric function can also be obtained if a dynamic collision frequency is calculated from optical data (here from KG results) and then used in a Mermin dielectric function~\cite{PhysRevLett.118.225001, 1367-2630-14-5-055020}. As can be seen in Fig. \ref{Diamond_fcc_eps_bse_tddft_rpa_x}, such a procedure leads to results different from both BSE and TDDFT, partly due to the different shape of KG in the optical limit and partly due to a different damping behaviour of the Mermin dielectric function with an increase in wavenumber. A difficulty that always arises when using the Mermin approach is the need to choose a charge state $Z$ i.e. the charge state of the ion, a choice that will have significant impact on the final result as the two curves for $q=0.76\ \AA^{-1} $ show. There is no generally valid procedure to extract the average charge state of an ion in the warm dense matter or high pressure solid range. As most models for ionization states work for hot, low density plasmas, the most promising seems to be the use of conductivity sum rule,

\begin{equation}
\centering
\int_{-\infty}^{+\infty} \frac{d \omega}{\pi}  \omega \Im[ \epsilon(q,\omega) ] = \omega_{p}^{2}
\end{equation} 

where $w_{p}$ is the plasma frequency and a split of the conductivity into Drude like free part and core part~\cite{PhysRevLett.118.225001}.       

Of paramount importance is also the imaginary part of the inverse dielectric function
as it can be used to compute further quantities like the dynamic structure factor or the stopping power that are directly experimentally accessible. In figures \ref{Diamond_fcc_eps_bse_tddft_rpa} \& \ref{Diamond_fcc_bse2}, the predictions of different models for different densities in the optical limit are shown. The main focus is on the plasmon peak, given by the zero of the real part of the dielectric function that lies in the energy range of weak damping, describing the collective excitation of valence electrons across the band gap. The variations in the location, height and width of the plasmon peak are far greater than in the imaginary part of the dielectric function. This is due to nonlinear effects introduced by the Kramers-Kronig relation between the real and imaginary parts of the dielectric function. Thus, the imaginary part of the inverse dielectric function is a necessary quantity in order to assess the quality of a theory in addition to the imaginary part of the dielectric function, i.e. the absorption. This holds in particular, as it is sensitive to different physics in a different energy range.

Similarly to the situation with the imaginary part, the plasmon peaks are located at the lowest energies in the BSE approximation. RPA and TDDFT predict similar plasmon locations at higher energies. Further, the width of the plasmon is smallest for the BSE and a lot greater for TDDFT/RPA. This means a much more stable quasi-particle is predicted within BSE. The plasmon peaks shown in the inset of Fig. \ref{Diamond_fcc_bse2} show the shift to higher energies as well as a broadening with increase of pressure.  

\begin{figure}[t]
\centering
\includegraphics[width=0.6\columnwidth]{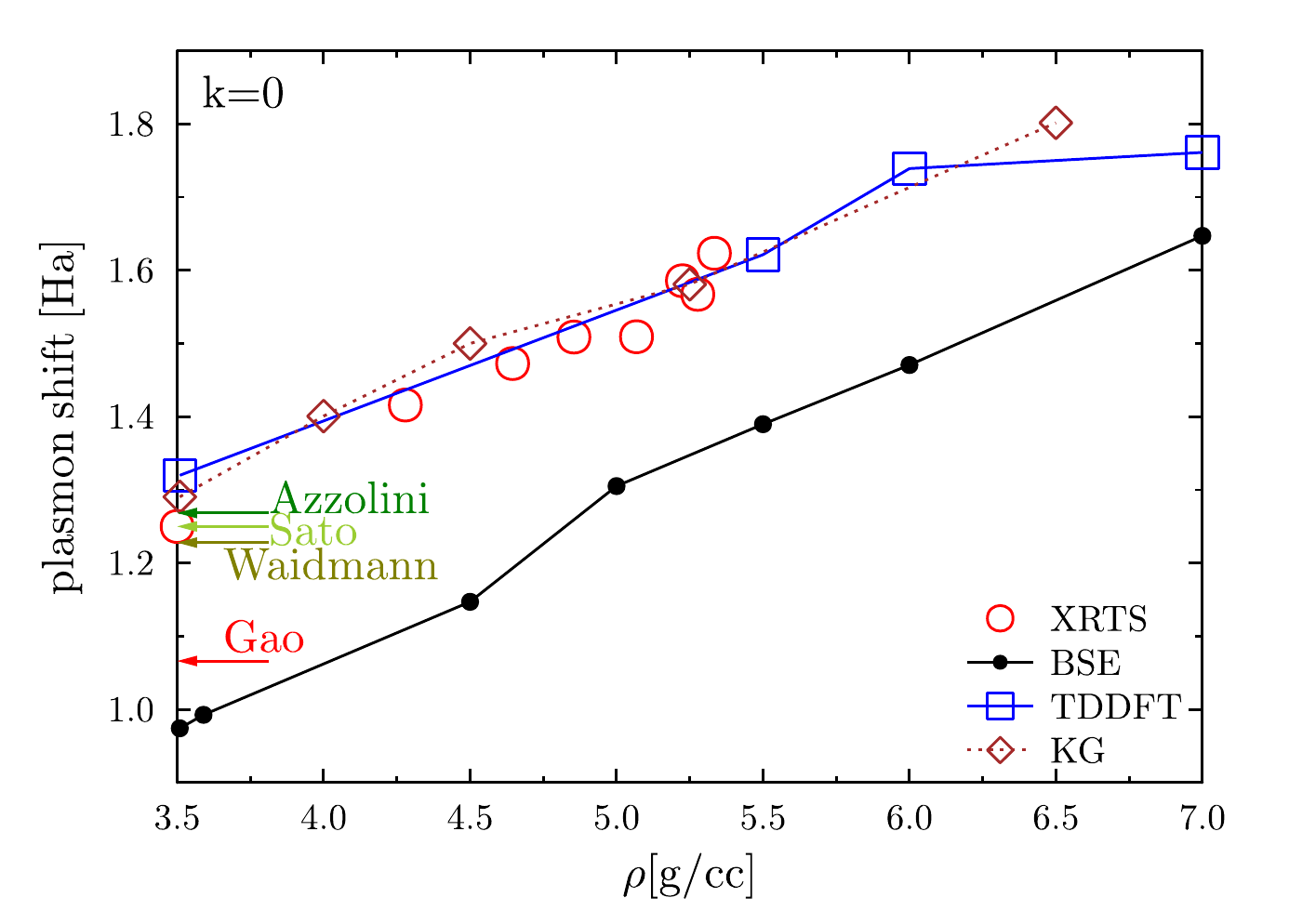}   
\caption{ \small{The position of the plasmon in diamond in the optical limit as function of the density. Experimental data obtained via x-ray Thompson scattering (XRTS) is taken from~\cite{osti_1241296}, electron-energy loss spectroscopy data by Sato \etal~\cite{sato_2012} and Waidmann \etal~\cite{PhysRevB.61.10149}, BSE calculations for ambient conditions by Gao~\cite{PSSB:PSSB201451197}, TDDFT calculations at ambient conditions by Azzolini \etal~\cite{azzolini_2017}. KG, BSE, and TDDFT results of this work.}}  
\label{plasmon_shift_density}
\end{figure}

In figure \ref{plasmon_shift_density}, we show the change of the location of the plasmon at $k=0$ with density as it can be obtained in experiment and theory. The slope, i.e., the increase of the plasmon peak energy with density, is predicted by all theories in agreement with the XRTS measurements. The absolute values of the plasmon positions is lowest using BSE, about $30\%$ lower than the TDDFT, RPA, KG, and available experimental results. Our KG and TDDFT results track the XRTS data nicely over the whole available experimental density range. Only for the highest densities where we reach pressures of up to $900\,$GPa, close to the proposed diamond to BC8 transition, differences between KG and TDDFT appear. The observed differences between theories at ambient diamond density are due to the implementation of scissor corrections in our TDDFT results to incorporate the correct band gap vs. an omission of scissor corrections in Azzolini \etal~\cite{azzolini_2017}.

\begin{figure}[t]
\centering
\includegraphics[width=0.6\columnwidth]{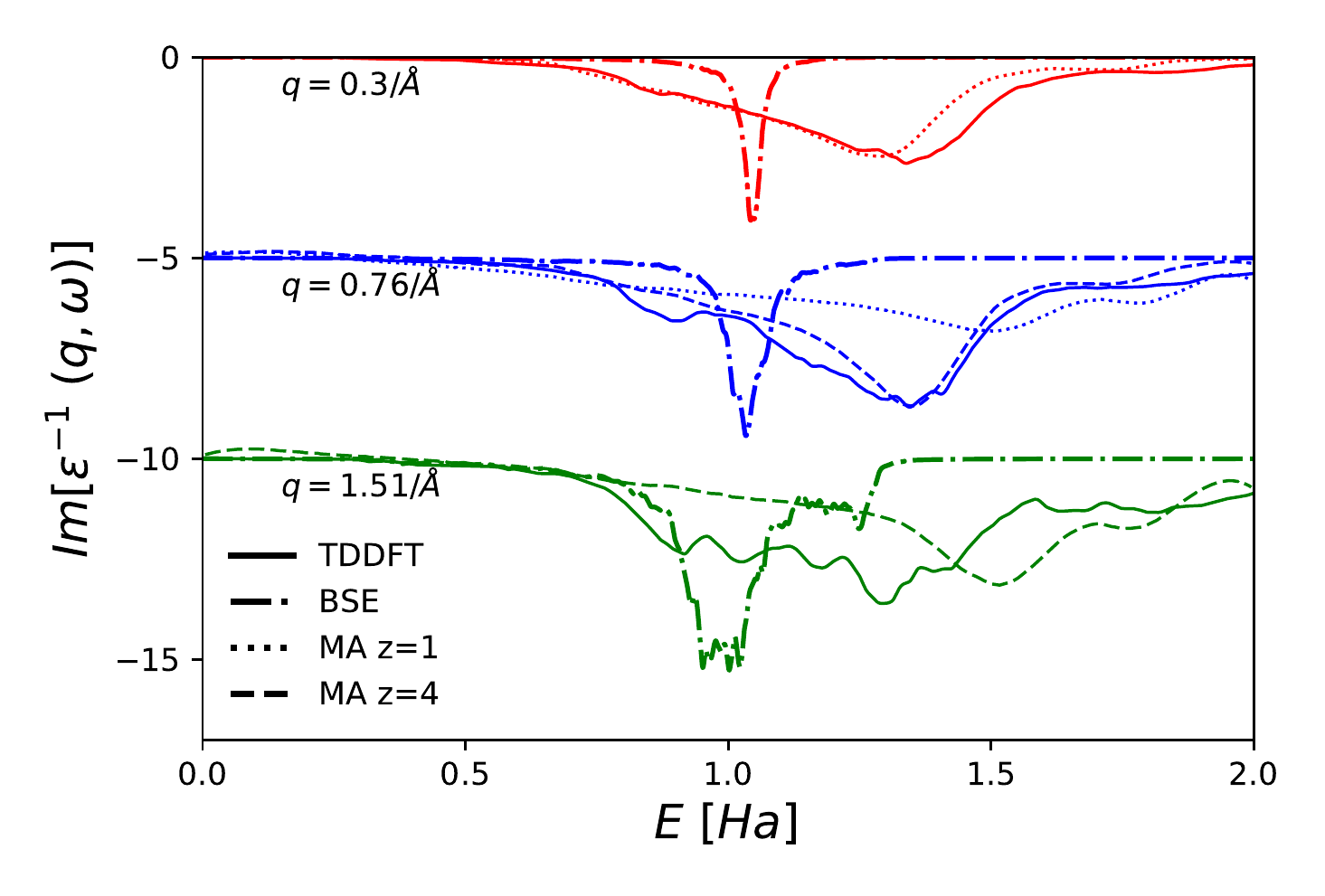}
\caption{ \small{The imaginary part of the inverse of the dielectric function for diamond at ambient density $\rho=3.509$ g/$cm^{3}$ using TDDFT, BSE and MA at finite $\vec{q}$. The charge state $Z=1$ is considered for $q=0.3/ \AA$ and $0.76/ \AA$ and $Z=4$ is considered for $q=0.76/ \AA$ and $1.51/ \AA$. } }  
\label{Diamond_mermin_tddft} 
\end{figure}   

Next we study the dispersion of the plasmon subject to a change in wavenumber. For this purpose, Fig. \ref{Diamond_mermin_tddft} displays the imaginary part of the inverse of the response function calculated using various approaches at finite \emph{q}. For small wavenumbers, one obviously observes a similar situation as in the optical limit. BSE predicts a too small excitation energy. TDDFT and the KG result extended using the Mermin approach predict nearly the same peak location but slightly different peak widths. Again, we point out the free parameter of the ion charge state entering the KG+Mermin formulation which is not known a priori~\cite{PhysRevE.81.026405}. For the highest value of $q=1.51\ \AA^{-1}$ the TDDFT and KG+Mermin results differ strongly in peak position and width indicating very different dispersion relations.  

\begin{figure}[t]
\centering
\includegraphics[width=0.6\columnwidth]{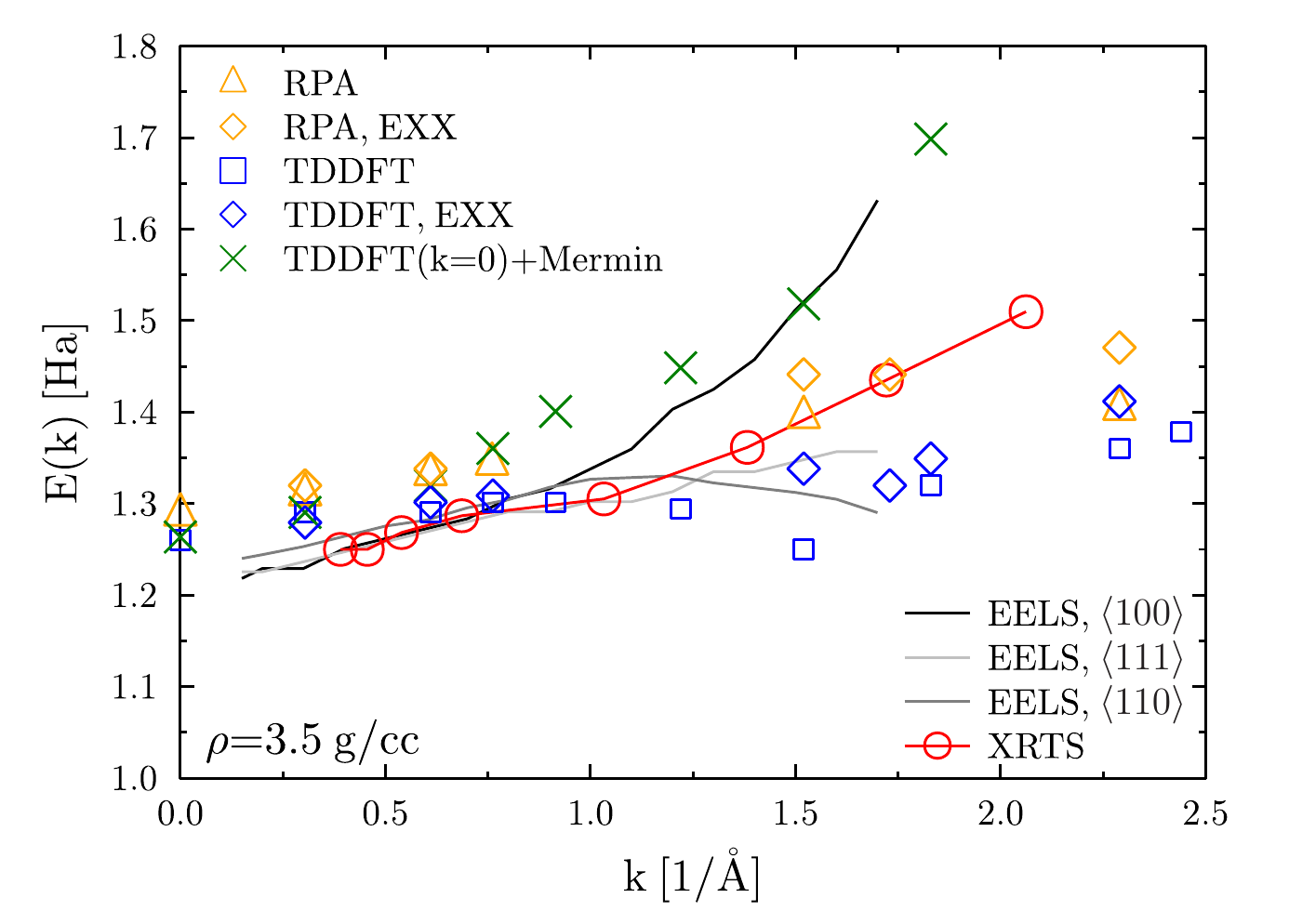}  
\caption{ \small{The position of the plasmon in diamond at ambient density as function of the wavenumber. Experimental data obtained via x-ray Thompson scattering (XRTS) is taken from~\cite{osti_1241296}, electron-energy loss spectroscopy (EELS) data by Waidmann \etal~\cite{PhysRevB.61.10149}. RPA, TDDFT and Mermin results of this work.} }
\label{Diamond_k_dispersion}
\end{figure}  

The change of the plasmon with wavenumber is shown in Fig. \ref{Diamond_k_dispersion}. We have omitted the BSE results, for which we already know the value in the optical limit to be too small. BSE predicts further no shift in the plasmon peak as function of the wavenumber within the error bars. Some experimental results in Fig. \ref{Diamond_k_dispersion} stem from electron loss spectroscopy, which is even able to resolve the slightly different dispersion relation in different lattice directions~\cite{PhysRevB.61.10149}. The XRTS method cannot do so, the results are basically an arithmetic mean of the EELS data showing a weak quadratic dispersion in \emph{q}~\cite{osti_1241296}. The TDDFT(or KG)+Mermin dispersion, using the DFT data from the optical limit, is by default quadratic as well, but the prefactor is too large giving blue shifted plasmon positions for large wavenumbers. For small wavenumbers, KG+Mermin (or TDDFT+Mermin) gives the same result as TDDFT. The best agreement with experiments over a wider wavenumber range can be reached when using first principle TDDFT or RPA methods that are explicitly capable of calculating the wavenumber dependence. We show results of different classes of approximations: with or without long range kernel (RPA vs. TDDFT) and with standard PBE-GGA exchange correlation functional versus exact exchange (EXX). Including local field corrections, i.e., the long range kernel $f_{xc}$, lowers the plasmon energies, compare the RPA and TDDFT curves. Both RPA and TDDFT show a change in plasmon energy with wavenumber that is smaller than the experimental values. A very flat plasmon dispersion similar to the current TDDFT results has also been reported by Azzolini \etal~\cite{azzolini_2017}. Improvements  with respect to the measurements are possible when using not a PBE-GGA exchange correlation functional but a more advanced exact exchange (EXX) functional. 

The best method overall appears to be TDDFT when using an advanced EXX functional. However, it seems that the TDDFT long range kernel has so far mostly been tested and benchmarked on absorption data in the optical limit. The present theoretical data show deficiencies for finite \emph{q} and higher energies. Areas where improvements are necessary can be found for instance in the $1 \ldots 2$ Ha energy range. 

\begin{figure}[t]
\centering
\includegraphics[width=0.6\columnwidth]{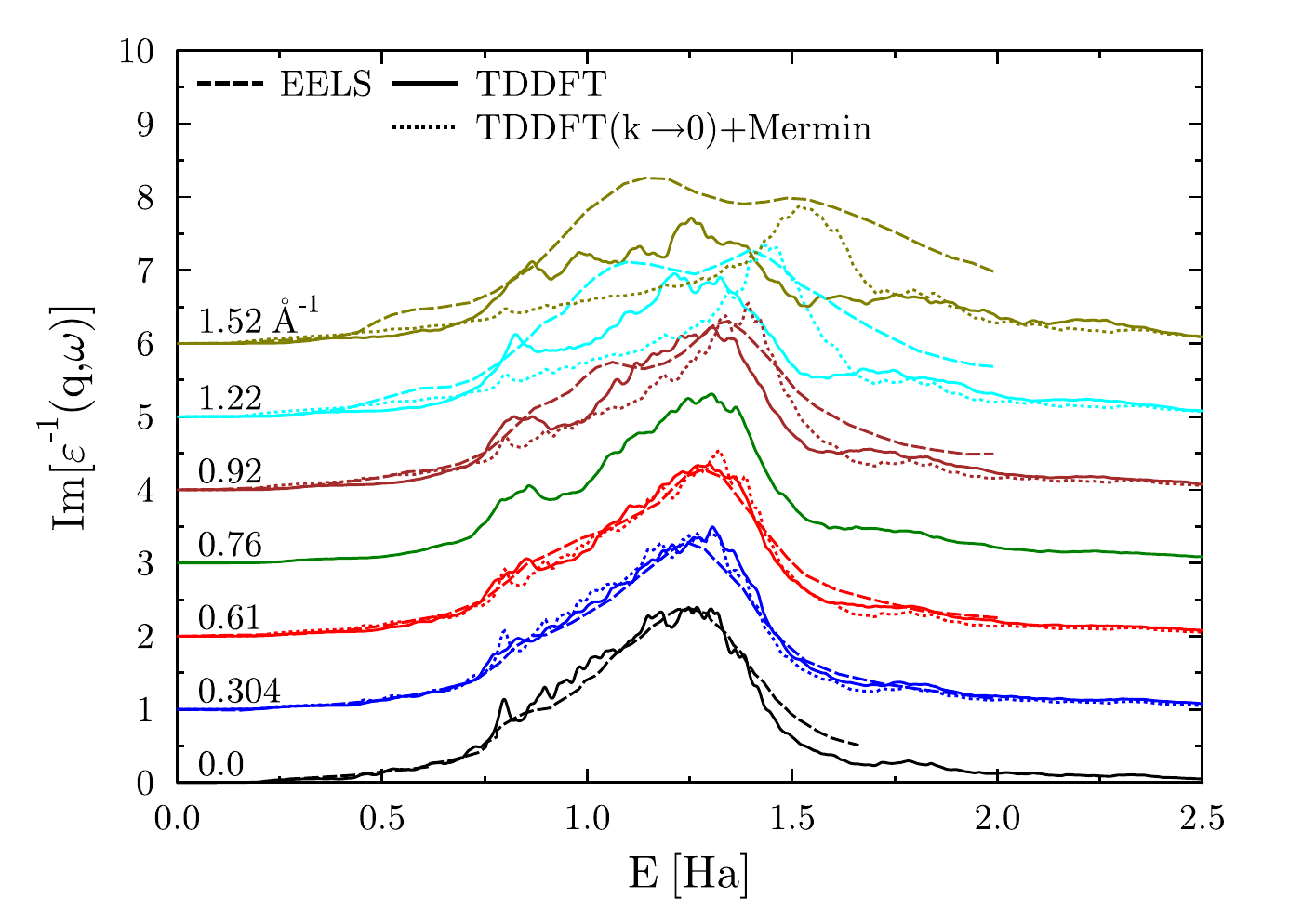}
\caption{ \small{The imaginary part of the inverse dielectric function for different wavenumbers for diamond at ambient conditions. Electron-energy loss spectroscopy (EELS) data by Waidmann \etal~\cite{PhysRevB.61.10149}, Daniels \etal~\cite{Daniels_1970}, and Raether~\cite{raether_2006}. TDDFT and Mermin results of this work.} }
\label{Diamond_k_full_tddft_exp} 
\end{figure}

Beside the location and dispersion of the plasmon peak, the general functional form of the imaginary part of the inverse dielectric function as function of energy and wavenumber is of interest. In Fig. \ref{Diamond_k_full_tddft_exp}, we compare experimental results with our supposedly best theoretical model. Again, for lower wavenumbers, the agreement is satisfactory. However, for the higher wavenumbers shown, we find that peaks and distribution of weight deviate between experiment and theory thus giving a more complete picture of the challenges for theory than just, e.g., the plasmon location.

\begin{figure}[t]
\centering
\includegraphics[width=0.6\columnwidth]{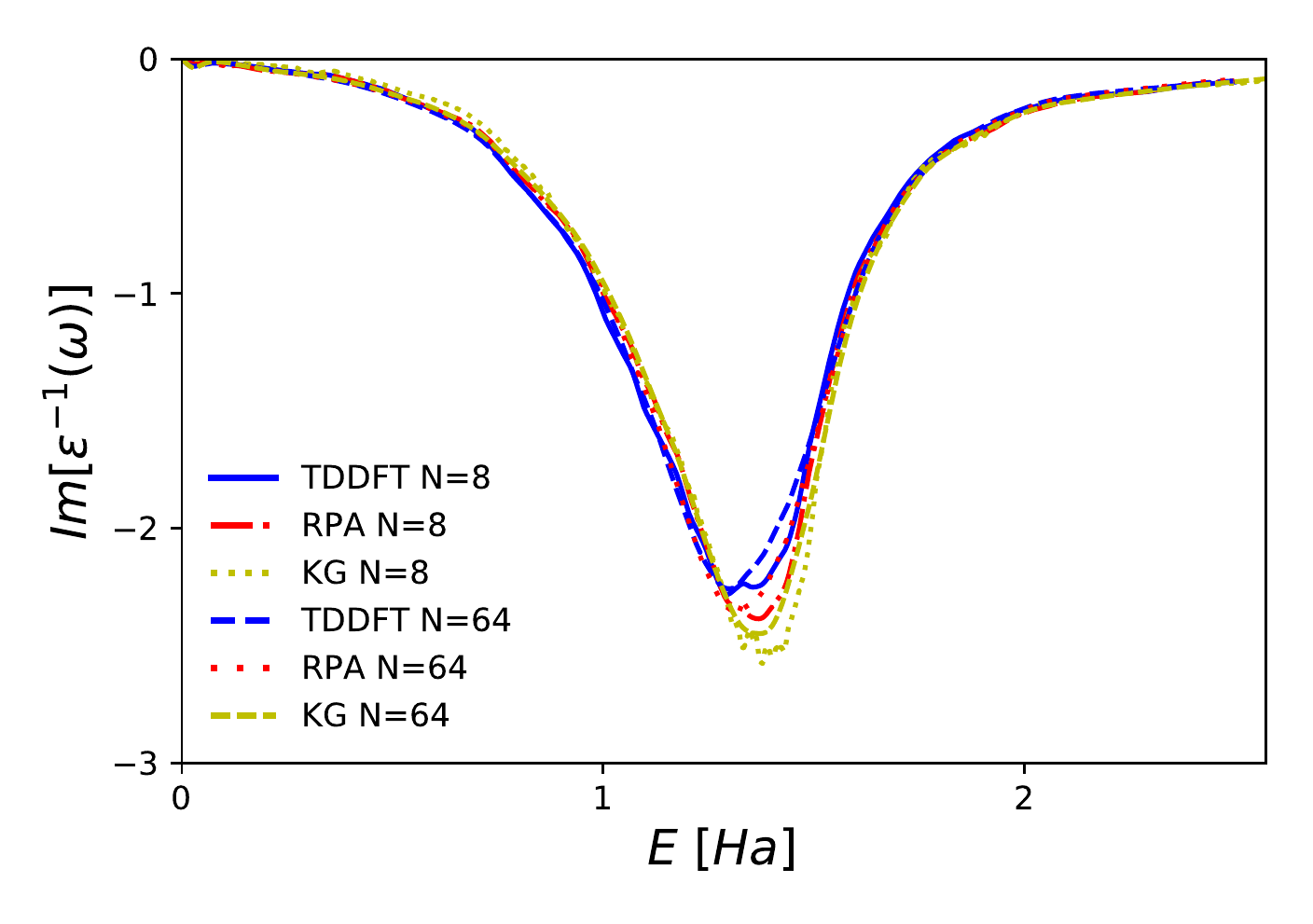} 
\caption{ \small{The imaginary part of the inverse dielectric function for diamond at warm dense matter conditions ($p=150$ GPa, $T=6000$ K) in the optical limit. } }
\label{Diamond_wdm}
\end{figure} 

A similar analysis as for a high pressure solid can be performed for a warm dense state of carbon as presented in Fig. \ref{Diamond_wdm}. It is apparent that it is essential to choose the system size as large as possible to capture all the possible excitations, in particular for the TDDFT \& RPA results that seem more dependent on the system size as the KG results. We provide additional results from TDDFT calculations concerning the influence of the xc-kernel and of smearing in \ref{app_tddft_xc}. Comparing the results of Figs. \ref{Diamond_fcc_bse2}, \ref{plasmon_shift_density} \&  \ref{Diamond_wdm}, we observe a down-shift of the plasmon energy with increase in temperature analogue to the closing of the band gap due to temperature effects. 
\par 
In general, evaluating all the information of this section, this emphasizes the need for TDDFT in calculating inelastic scattering spectra since it compares best with known experimental results and has less computational demands than BSE~\cite{PhysRevLett.107.186401, PhysRevB.69.155112, Sharma_2012}. In particular, TDDFT can also be used on DFT-MD snapshots in the warm dense matter regime (similar to the workflow when using the KG approach, but without the need to guess a charge state), where the theoretical  $S(q,\omega)\propto$ Im[$\epsilon^{-1}(q,\omega)]$ is used in experiments to fit the XRTS signal, see Kraus \etal for diamond at 150 GPa and 5000 K~\cite{Kraus2017}.


\subsection{Lonsdaleite}

\par
Lonsdaleite is an allotrope of carbon with the spacegroup P$6_{3}$/mmc. The lattice parameters are $a =b \neq c$, $\alpha=\beta=90^{0},\ \gamma=120^{0}$. The unit cell volume is $ (\sqrt{3}/2)  a^{2} c $ with 4 atoms per unit cell. The basis vectors in lattice coordinates are given by the 4f Wyckoff positions $(\frac{1}{3},\frac{2}{3},z_{1})$, $(\frac{2}{3},\frac{1}{3},\frac{1}{2}+z_{1})$, $(\frac{2}{3},\frac{1}{3},-z_{1})$, $(\frac{1}{3},\frac{2}{3},\frac{1}{2}-z_{1})$ with the internal parameter $z_{1}$~\cite{MEHL2017S1}. 


The DFT calculations were performed using the elk code for 16 bands on a $16\times16\times16$ \emph{k}-point mesh using PBE-GGA functional with Broyden mixing~\cite{PhysRevLett.77.3865}. The muffin-tin radius was adjusted ranging from $0.55$ to $0.75$ \AA\hspace{1pt} to account for the different pressure ranges. The equilibrium lattice constants using the Vinet equation are calculated to be $a=2.524$ \AA, $c=4.128$ \AA \hspace{1 pt} with $z_{1}=0.0625$ which are in excellent agreement with the experimental values (a$=$2.52 \AA, c$=$4.12 \AA)~\cite{doi:10.1063/1.1841236} and theoretical results~\cite{PhysRevB.36.3373, PhysRevB.35.5856, PhysRevMaterials.1.054603}. The ideal values for the hexagonal structure are given by $z_{1}=\frac{1}{16}$ with $c/a=\sqrt{8/3}$. The ideal value of $z_{1}=0.0625$ for $c/a=1.635$ is favorable up to $600$ GPa based on the enthalpy changes for the aforementioned equilibrium lattice parameters. The methodology to obtain these parameters are discussed in \ref{lons_appendix}.

\begin{figure}[t]
\centering
\includegraphics[width=0.6\columnwidth]{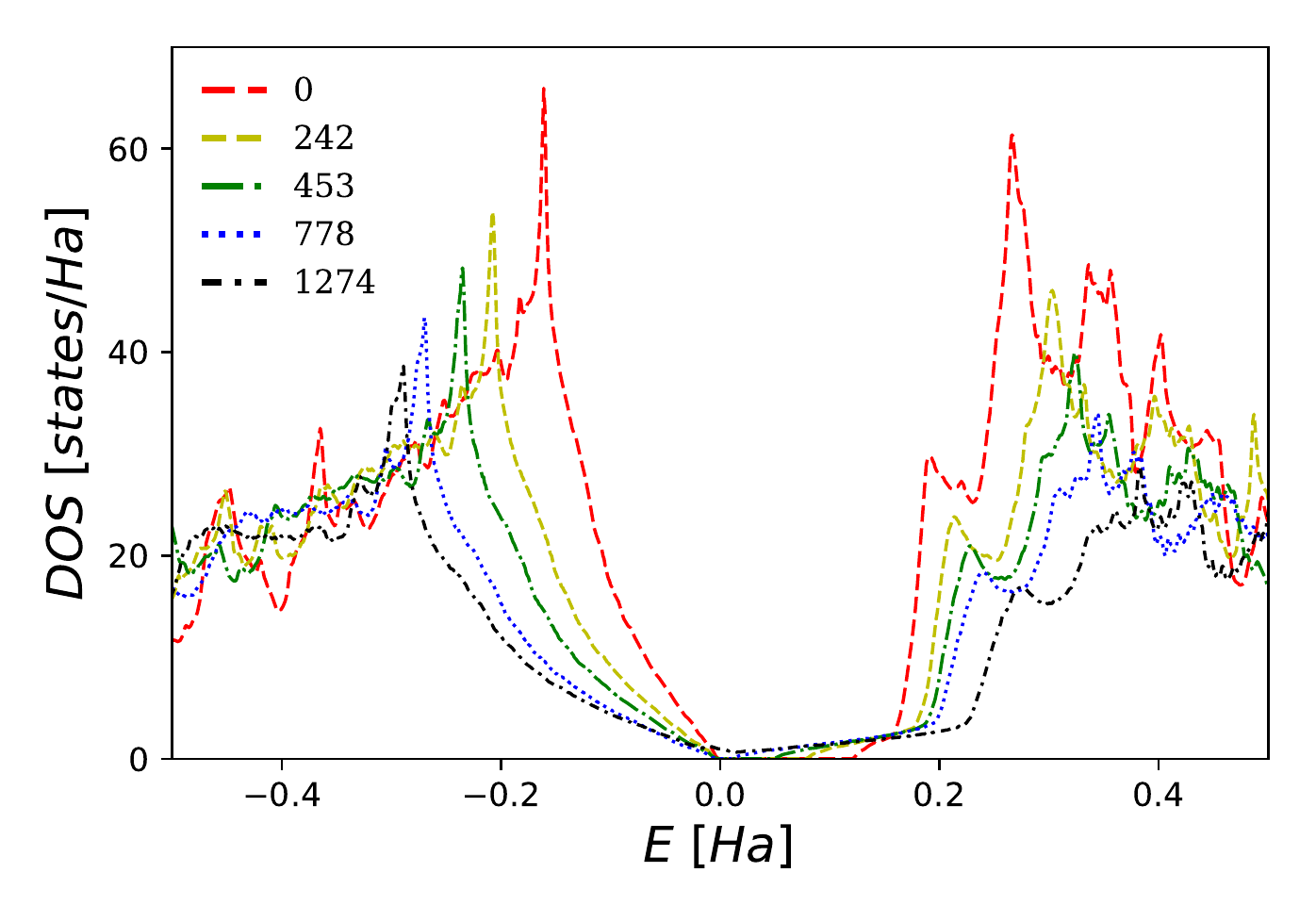}
\caption{ \small{The density of states for lonsdaleite as function of pressure as obtained from DFT using the all-electron full potential code elk. The pressures are in GPa and the valence band maximum is adjusted to zero.} }
\label{lonsdaleite_dos_compare}
\end{figure}

\begin{figure}[t]
\centering
\includegraphics[width=0.6\columnwidth]{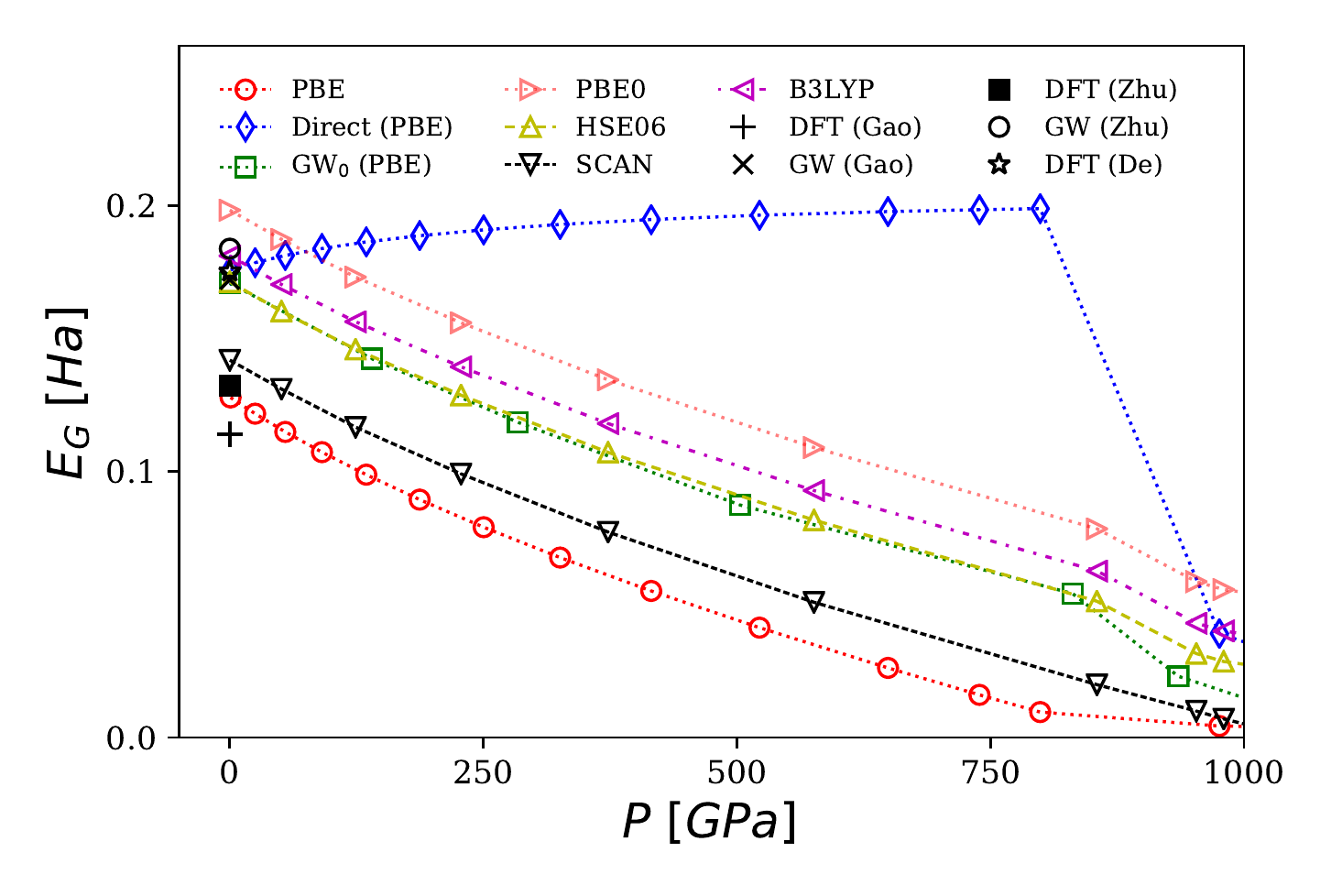}
\caption{ \small{Lonsdaleite band gap as function of pressure using DFT and GW. We also show the direct band gap calculated using PBE. Gao uses LDA and the routines implemented in abinit~\cite{PSSB:PSSB201451197}; Zhu \etal use PBE-GGA within VASP~\cite{PhysRevB.83.193410}; De \etal results are based on an empirical pseudopotential method~\cite{0953-8984-26-4-045801}. The density range covers $3.5$ g/cc to $6.9$ g/cc.}  
} 
\label{Lonsdaleite_dft_gw_bse} 
\end{figure}

The electronic density of states is shown in Fig. \ref{lonsdaleite_dos_compare}. Contrary to cubic diamond, the band gap closes with pressure increase. To account for the corrections of the eigenvalues, we use the GW approximation implemented in the VASP code with a 16$\times$16$\times$16 \emph{k}-point mesh centered around the Gamma point for $64$ bands, and a PBE-GGA exchange-correlation functional. The kinetic energy cutoff for the wavefunctions was set to $40$ Ha. We used the single shot GW$_0$ approach ignoring the off-diagonal elements of the self energy. The lowest band gap is indirect just like for the cubic diamond phase but across the points $\Gamma \rightarrow K$~\cite{PhysRevB.41.3048}. The band gap reduces with increasing pressure as shown in Fig. \ref{Lonsdaleite_dft_gw_bse}. The GW and the PBE band gaps are in good agreement with the available theoretical values for zero pressure~\cite{PSSB:PSSB201451197, PhysRevB.83.193410, 0953-8984-26-4-045801}. Again, HSE06 seems the best of the advanced xc-functionals in comparison with GW, even though B3LYP is close. 
  
\par  
The calculations for the dielectric function based on TDDFT were performed using the Bootstrap kernel on a $12 \times 12 \times 12$ \emph{k}-point mesh and $24$ empty states. The BSE calculations were done on a lower $8 \times 8 \times 8$ \emph{k}-point mesh and 16 empty states. So far, no known experimental band gaps are reported for lonsdaleite. With a large band gap, the screened Coulomb interaction is strong and the excitonic effects are prominent at lower pressures. With increasing pressure, contrary to the case of diamond, the band gap is reduced until it finally closes in the vicinity of $1000$ GPa.

\begin{figure}[t] 
\centering
\includegraphics[width=0.6\columnwidth]{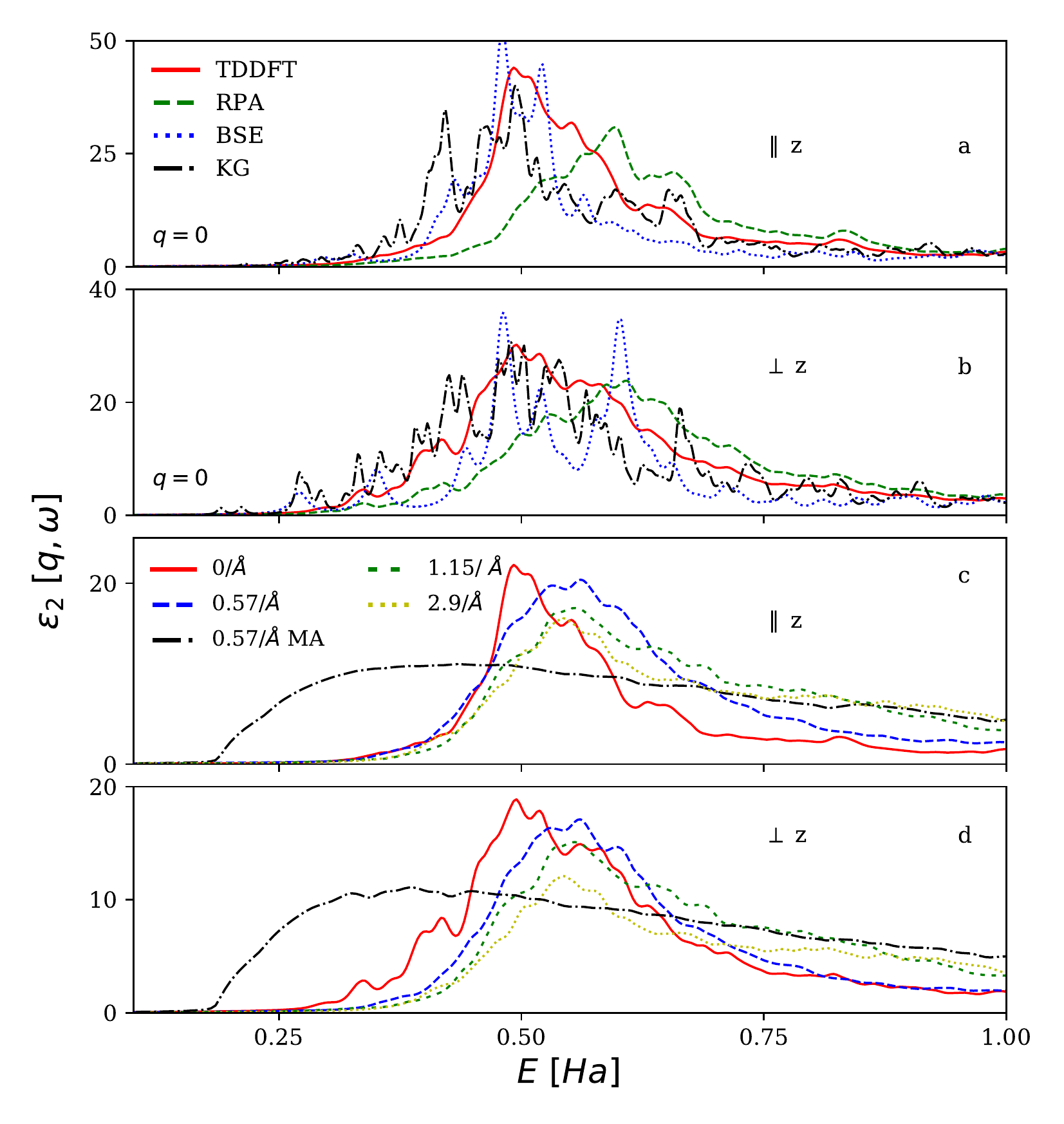}  
\caption{ \small{The imaginary part of the dielectric response function in panels a) and b) for lonsdaleite at 242 GPa using various approaches at $\vec{q}$=0. We compare TDDFT to KG+Mermin (MA)
at finite $\vec{q}$ in panels c) and d).  }
}  
\label{Lonsdaleite_242GPa_q_im_eps}
\end{figure} 
  
Figure \ref{Lonsdaleite_242GPa_q_im_eps} shows the imaginary part of the response function using various approaches along two orientations for the hexagonal lattice structure at $\vec{q}=0$ and using TDDFT at finite \emph{q}. The spectrum in the optical limit is well represented by BSE, and the TDDFT spectra is blue-shifted at a lower oscillator strength.  In this case, TDDFT doesn't quite resolve the prominent peaks in the BSE spectra along both the directions.
The TDDFT maximum is located at the average of the twin peaks of the BSE result for the perpendicular case and closer to the first peak in the parallel case. As in the case for diamond, RPA predicts the maximum of the imaginary part of the dielectric function at higher energies than BSE or TDDFT for the case $\parallel$ to $z$. On the other hand, the KG calculation shows a red shift compared to BSE and TDDFT. For the case $\perp$ to the z axis, all difference are mitigated.  

In panels c) and d) of Fig. \ref{Lonsdaleite_242GPa_q_im_eps}, we show finite-\emph{q} results from TDDFT together with results from the Mermin response function using a DFT(KG) based collision frequency. For $||\vec{q}||=0.57\ \AA^{-1} $, we choose an ionization degree $z=1$ as the fitting parameter. At $\vec{q} \neq 0$, the Mermin+DFT spectrum is almost uniform along both the directions in stark contrast to the TDDFT curves, as can be seen in Fig. \ref{Lonsdaleite_242GPa_q_im_eps}. In addition, we observe that the differences between the two directions parallel and orthogonal to $z$, respectively, as apparent in the optical limit (panels a) and b) in Fig. \ref{Lonsdaleite_242GPa_q_im_eps}, vanish for the finite \emph{q} values as shown in panels c) and d).

\begin{figure}[t]
\centering 
\includegraphics[width=0.6\columnwidth]{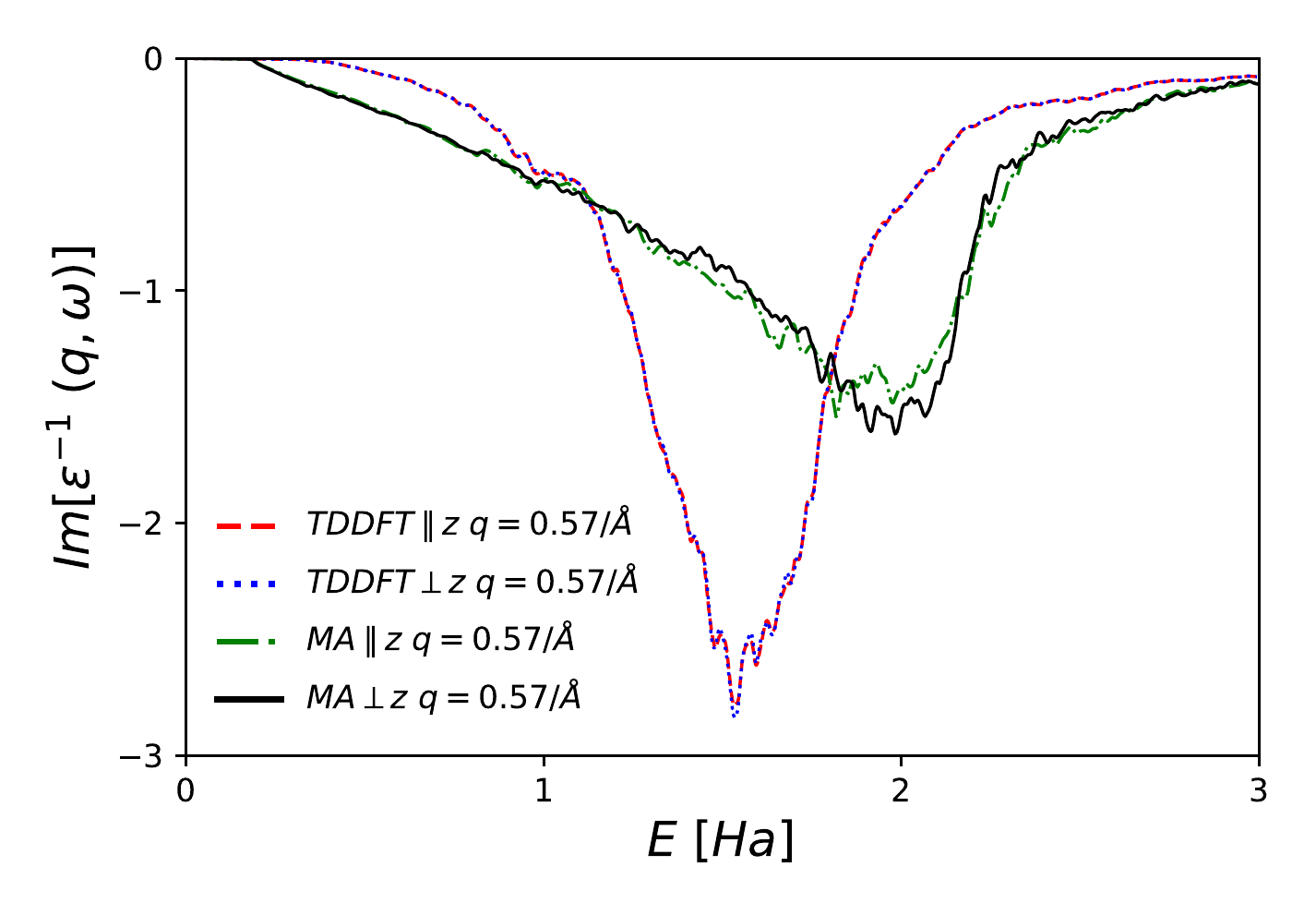}
\caption{ \small{The imaginary part of the inverse of the dielectric function for lonsdaleite at 242 GPa using MA and TDDFT at $||\vec{q}||=0.57 \AA^{-1}$. } }
\label{Lonsdaleite_242GPa_q_im_inv_eps}
\end{figure}  

In Fig. \ref{Lonsdaleite_242GPa_q_im_inv_eps}, the imaginary part of the inverse dielectric function for lonsdaleite at a pressure of $p=242$ GPa is presented. 
The difference in plasmon peak positions between TDDFT and Mermin+DFT is around $0.3-0.4$ Ha, which is larger than in the cubic diamond case. Further, the Mermin+DFT result has different structures near the peak parallel and perpendicular to $z$ compared to TDDFT in Fig. \ref{Lonsdaleite_242GPa_q_im_inv_eps}.  

\begin{figure}[H] 
\centering
\includegraphics[width=0.6\columnwidth]{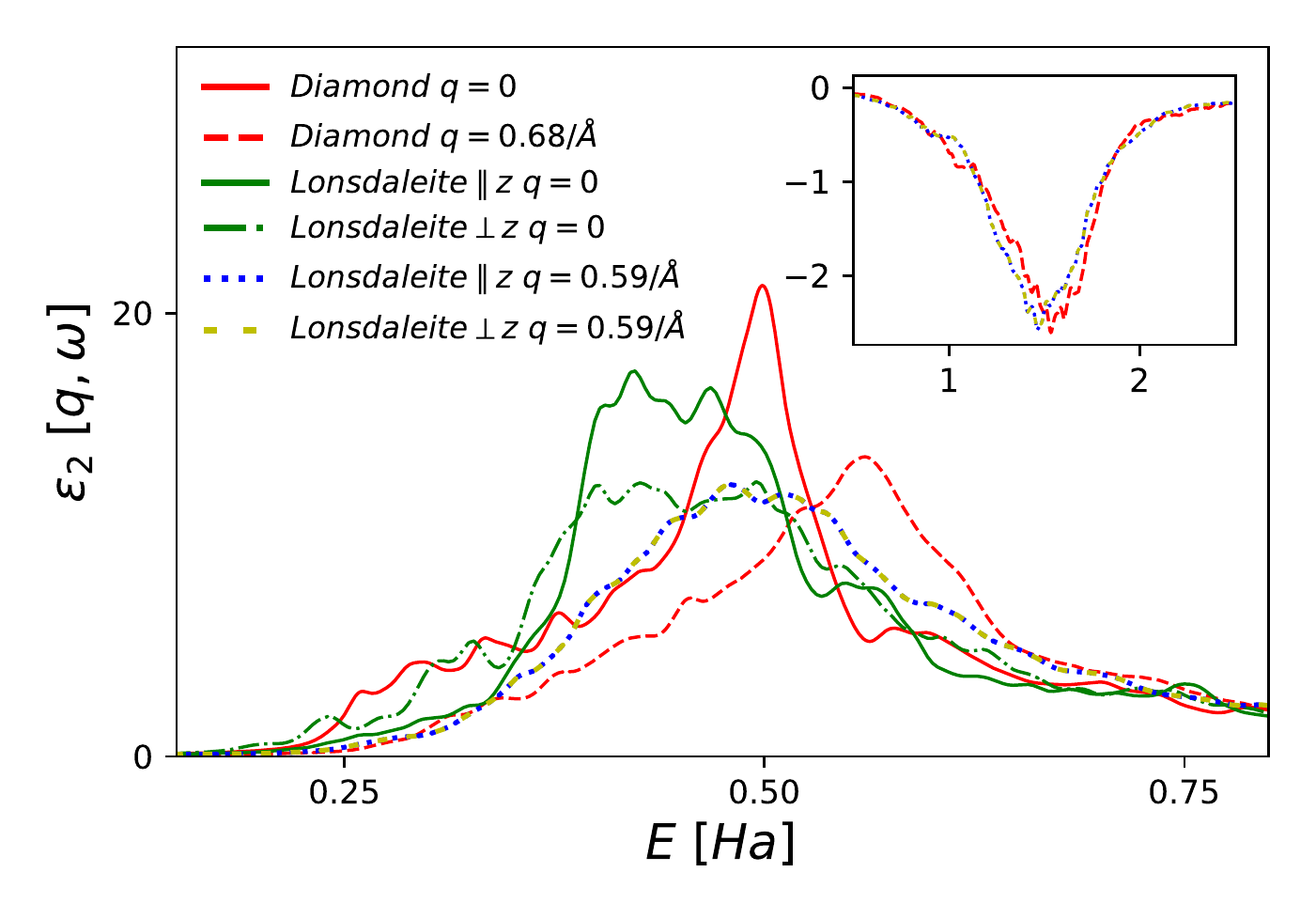}
\caption{ \small{Comparison of the imaginary part of the TDDFT dielectric function for diamond and lonsdaleite at 292 GPa in the optical limit and at finite \emph{q}. The inset panel shows the imaginary part of the inverse of the dielectric function at $||\vec{q}||=0.68\ \AA^{-1} $ for diamond and $||\vec{q}||=0.59\ \AA^{-1}$ for lonsdaleite (red--diamond, blue/yellow--lonsdaleite).} }
\label{fcc_lons_eps_compare_292GPa}
\end{figure}  
   
In an x-ray scattering experiment, it would be of advantage to not only distinguish the diamond and lonsdaleite phases in the x-ray diffraction (elastic) signal, but also to find characteristic traits in the inelastic spectrum. Figure \ref{fcc_lons_eps_compare_292GPa} shows that it should indeed be possible to distinguish lonsdaleite from diamond at high pressure conditions either in the absorption spectrum or in the inelastic scattering spectrum by the characteristic positions of the respective absorption and plasmon peaks. 


\subsection{BC8}
BC8 has a body-centered cubic structure with the spacegroup Ia$\bar{3}$ consisting of $8$ atoms per unit cell. The unit cell volume is $a_{0}^{3}/2$ where $a_{0}$ is the lattice constant. The lattice parameters are $a=b=c$, $\alpha=\beta=\gamma=90^{0}$~\cite{PhysRevB.89.224109, PhysRevLett.118.146601}. The basis vectors in lattice coordinates are given by the 16c Wyckoff positions $(2x_{1},2x_{1},x_{1})$, $(\frac{1}{2},0,\frac{1}{2}-2x_{1})$, $(0,\frac{1}{2}-2x_{1},\frac{1}{2})$, $(\frac{1}{2}-2x_{1},\frac{1}{2},0)$, $(-2x_{1},-2x_{1},-2x_{1})$, $(\frac{1}{2},0,\frac{1}{2}+2x_{1})$, $(0,\frac{1}{2}+2x_{1},\frac{1}{2})$, $(\frac{1}{2}+2x_{1},\frac{1}{2},0)$ with $x_{1}$ as a parameter~\cite{MEHL2017S1}. 


The equilibrium value $x_{1}=0.0935$ compares favorably to the experimental value, $0.1003 \pm 0.0008 $ widely used for the bc8 phase of silicon. This value is also used for the DFT simulations of the bc8 carbon phase~\cite{ClarkStewart,Crain_1994,Kasper_1964}. The enthalpy changes up to 2500 GPa are negligible among the $x_{1}$ values considered and hence we consider the parameter $x_{1}=0.1003$ for all our calculations, see \ref{bc8_appendix}. Our equilibrium lattice constant a$_{0}$=4.437 \AA \hspace{1 pt}  lies within the range of the values 4.425 - 4.477 \AA \hspace{1 pt} obtained by Z. Li and Crain \etal~\cite{PhysRevB.91.214106, Crain_1994}.  

\begin{figure}[htp] 
\centering  
\includegraphics[width=0.6\columnwidth]{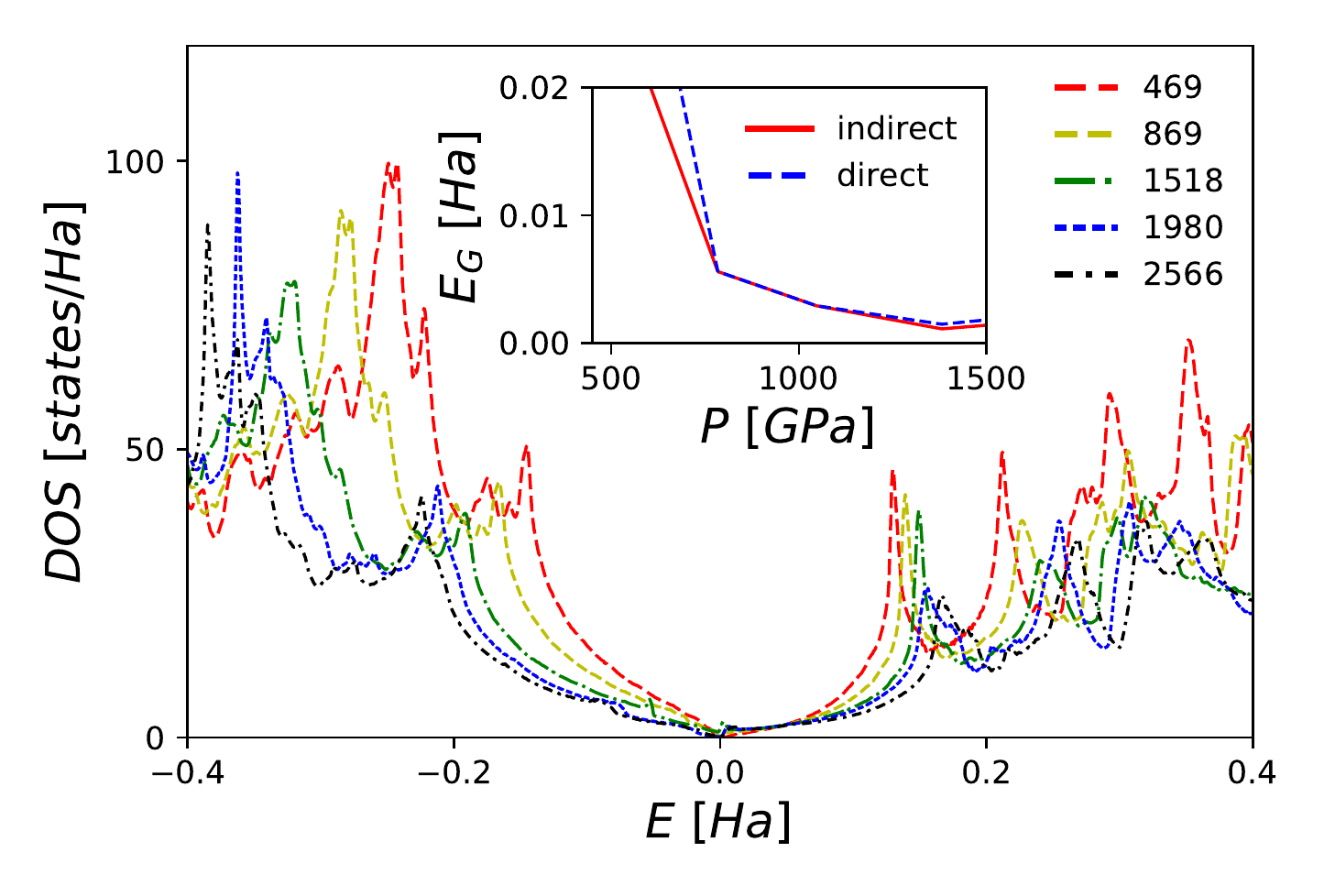}
\caption{ \small{The density of states for bc8 as function of pressure obtained from DFT using the all-electron full potential code elk. The inset plot shows the direct and indirect band gaps calculated using the PBE functional in elk. The pressures indicated are in GPa and the valence band maximum is adjusted to zero.} }
\label{bc8_dos_elk}  
\end{figure} 

In Fig. \ref{bc8_dos_elk}, the electronic density of states is shown as calculated using a PBE-GGA exchange-correlation functional in elk on a $16\times16\times16$ \emph{k}-point mesh for $32$ bands. The LDA (GW) bandgaps at zero pressure of $3.58$ and $2.7 \ (3.5)$ eV, respectively, reported by Z. Li and Zhu \etal are large compared to our GGA (GW) values $0.9\  (1.94)$ eV, which are comparable to early calculations by Johnston \etal~\cite{PhysRevB.91.214106, PhysRevB.83.193410,Jonston_1989}.
Correa \etal obtained a GGA bandgap of approximately 0.40 eV near the phase transition boundary from diamond at T=0 which may be compared to our result of 0.15 eV~\cite{Correa_2006}. The band gap reduces with increasing pressure and closes in the vicinity of 2900 GPa, where the simple cubic structure is the thermodynamically stable phase~\cite{PhysRevLett.108.045704}. 

\begin{figure}[t]
\centering
\includegraphics[width=0.6\columnwidth]{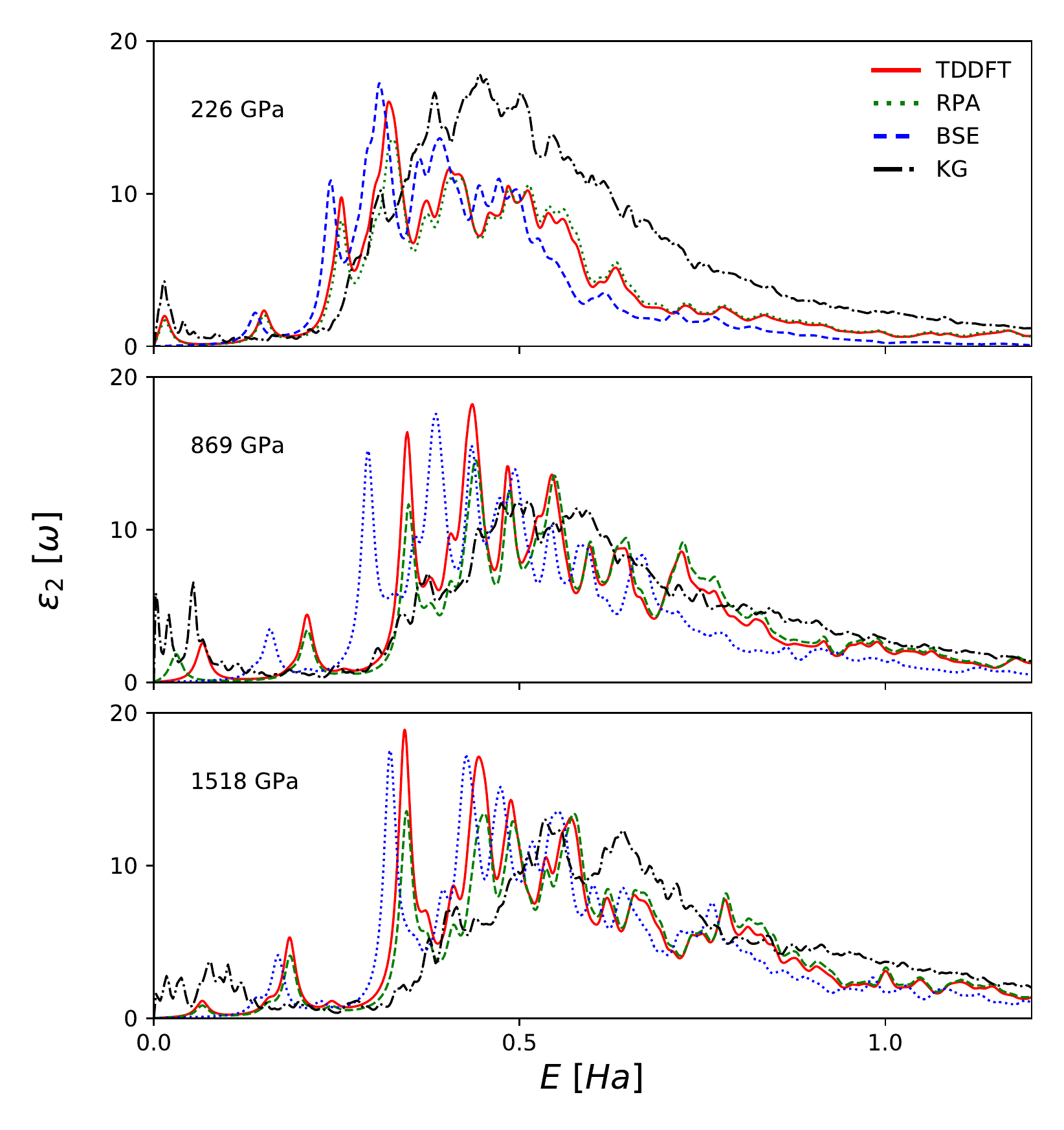}
\caption{ \small{The imaginary part of the dielectric function of bc8 in the optical limit for various pressures using BSE, TDDFT, RPA and Kubo-Greenwood formula. 
} }
\label{bc8_elk_TDDFT} 
\end{figure} 

The absorption spectrum is obtained using TDDFT with the bootstrap kernel for a $4\times4\times4$ \emph{k}-point mesh and $28$ empty states. The BSE calculations are also done on a similar $4 \times 4 \times 4$ \emph{k}-point mesh and 18 empty states. Due to the larger basis set size of BC8 compared to diamond, a similar \emph{k}-point mesh as in diamond is unfeasible. However, due to the close proximity of the band edges for small pressures, a smaller \emph{k}-point mesh is sufficient for convergence, see Fig. \ref{bc8_elk_TDDFT}. There is close resemblance of RPA, TDDFT and BSE spectra albeit with different peak locations for the different pressures. The KG-formula has a significant absorption peak near $\omega=0$ due to the nature of semi-metallicity from DFT prediction which is clearly absent from BSE spectra even at higher pressures. The KG result also show a shift of the main absorption peak to higher energies than found with RPA, TDDFT, and BSE. The overall agreement of BSE, RPA, and TDDFT is quite remarkable for BC8, especially in hindsight after having analyzed the diamond and lonsdaleite phases.
 
\begin{figure}[H] 
\centering  
\includegraphics[width=0.6\columnwidth]{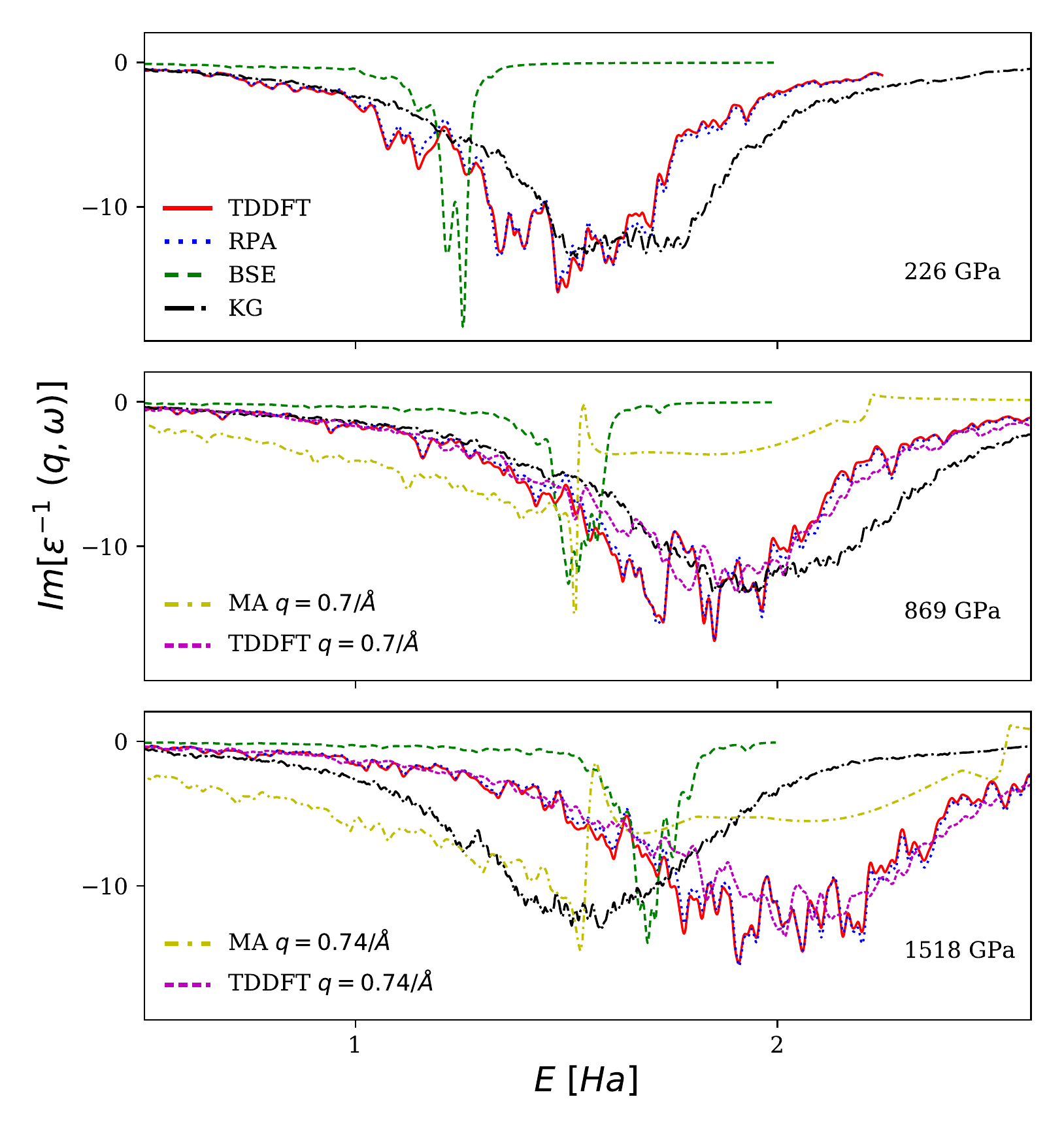}
\caption{ \small{The imaginary part of the inverse of the dielectric function of bc8 at various pressures using BSE, TDDFT, RPA and the Kubo-Greenwood formula. The TDDFT, RPA, and KG are scaled fivefold and the finite-\emph{q} MA curve is scaled tenfold. All curves in the optical limit except the specially marked MA and TDDFT curves. } }  
\label{bc8_elk_TDDFT_im_inv_eps}  
\end{figure}

Figure \ref{bc8_elk_TDDFT_im_inv_eps} shows the collective oscillations via the imaginary part of the inverse dielectric function obtained using various approaches. The TDDFT and RPA spectra do not show the twin peaks of the BSE and predict a very broad  feature instead. The KG-formula produces a broad peak at a different location.  Whereas the plasmon peak in BSE and TDDFT shifts to higher energies for higher pressures, the KG result shows the opposite behaviour for the case of the highest pressure. As seen earlier for diamond, the BSE curve predicts the most stable plasmon excitation (rather narrow peak with largest magnitude).

For the cases of 869 GPa and 1518 GPa, we also provide the results for finite wave numbers and a comparison between the MA approach and TDDFT. The collision frequency used in MA is taken from the optical limit of TDDFT. The TDDFT results in the optical limit and for finite wavenumbers are very similar owing to the small change in wavenumber. However, the MA approach shows drastically different behavior with the peak location at smaller energy and the magnitude reduced.

\section{Conclusions}

\par
We have calculated the electronic density of states, the band gap, and the dielectric response function as a function of pressure for various phases of carbon, namely diamond, lonsdaleite, and bc8. In particular, we were interested in the wavenumber dependence of the dielectric function in order to provide high quality predictions for the dynamic structure and XRTS signals of these phases under high pressure. We also calculated the dynamic structure for warm dense matter states of carbon.

The theory including the most advanced approximations is BSE which therefore was expected to provide the quantitatively best results. However, it is clear that we are at the limit of what is currently possible in terms of computer resources with BSE.
In this light, it is fortunate that TDDFT can reproduce the excitonic effects of the BSE reasonably well at a fraction of the computational cost and also at finite $\vec{q}$. We have further analyzed how the standard KG approach and its extension to finite wavenumbers via the MA theory compares to BSE and TDDFT. We investigated the influence of different xc-functionals on the band gaps and eigenvalues and thus on the response functions.

Available experimental data for diamond have been compared to the results of the simulations. For lonsdaleite and the bc8 phase, experimental data under high pressure are very sparse and only comparisons with different theoretical methods were possible. In the case of diamond, where the band gap increases with pressure, the HSE06 xc-potential gave the best agreement with GW calculations of the band gap. The change in the location of the optical plasmon with density was best reproduced by KG or TDDFT. The experimental dispersion of the plasmon at ambient conditions compares best with a TDDFT calculation without scissor correction and the EXX potential. Similarly, TDDFT can reasonably well give the functional form of the imaginary part of the inverse dielectric function for small wavenumbers but fails to reproduce the peak structure for higher wavenumbers as obtained from EELS measurements. For warm dense matter conditions, TDDFT and RPA give different results to KG, and particular care is needed to eliminate finite size effects. Moreover, it seems that for the case of diamond, the approximations inherent in the used BSE solver are not be justified as this method does not compare well to experimental results.

Lonsdaleite shows a decrease in band gap with increasing pressure and HSE06 seems again best when compared to GW calculations. The differences in the dielectric response between the directions parallel and perpendicular to the z-axis, that are apparent in the optical limit, seem to vanish for finite wavenumbers. We found substantial differences between MA and TDDFT for the imaginary part of the inverse dielectric function at high pressure. Based on our TDDFT results, it might be possible to distinguish diamond and lonsdaleite based on their XRTS signal in the warm dense matter range.

We performed an accurate analysis of the parameter $x_1$ of the bc8 lattice structure for a wide pressure range and found deviations of our DFT/GW bandgaps to predicted values for ambient conditions and for high pressures. Similarly to the diamond case, BSE gives plasmon positions at lower energies than all the other methods. Whereas BSE, RPA, and TDDFT show a continuous increase of the plasmon energy with pressure, the KG method is different. The approach using TDDFT to compute the inelastic-scattering spectra should be a viable tool for experiments involving carbon and carbon bearing mixtures e.g. with attention to the formation of diamond. Our results can provide a theoretical reference for future experiments on band gaps and optical properties for the various phases of carbon at high pressures. 

\ack
Computations were performed on a Bull Cluster at the Center for Information Services and High Performance Computing (ZIH) at TU Dresden. We would like to thank the ZIH for its support and generous allocations of computer time.

\section*{Appendix}

\appendix

\section{Pressure calculations, lattice parameters, band gaps}
\label{app_eos}

\begin{figure}[H]
\centering 
\includegraphics[width=0.6\columnwidth]{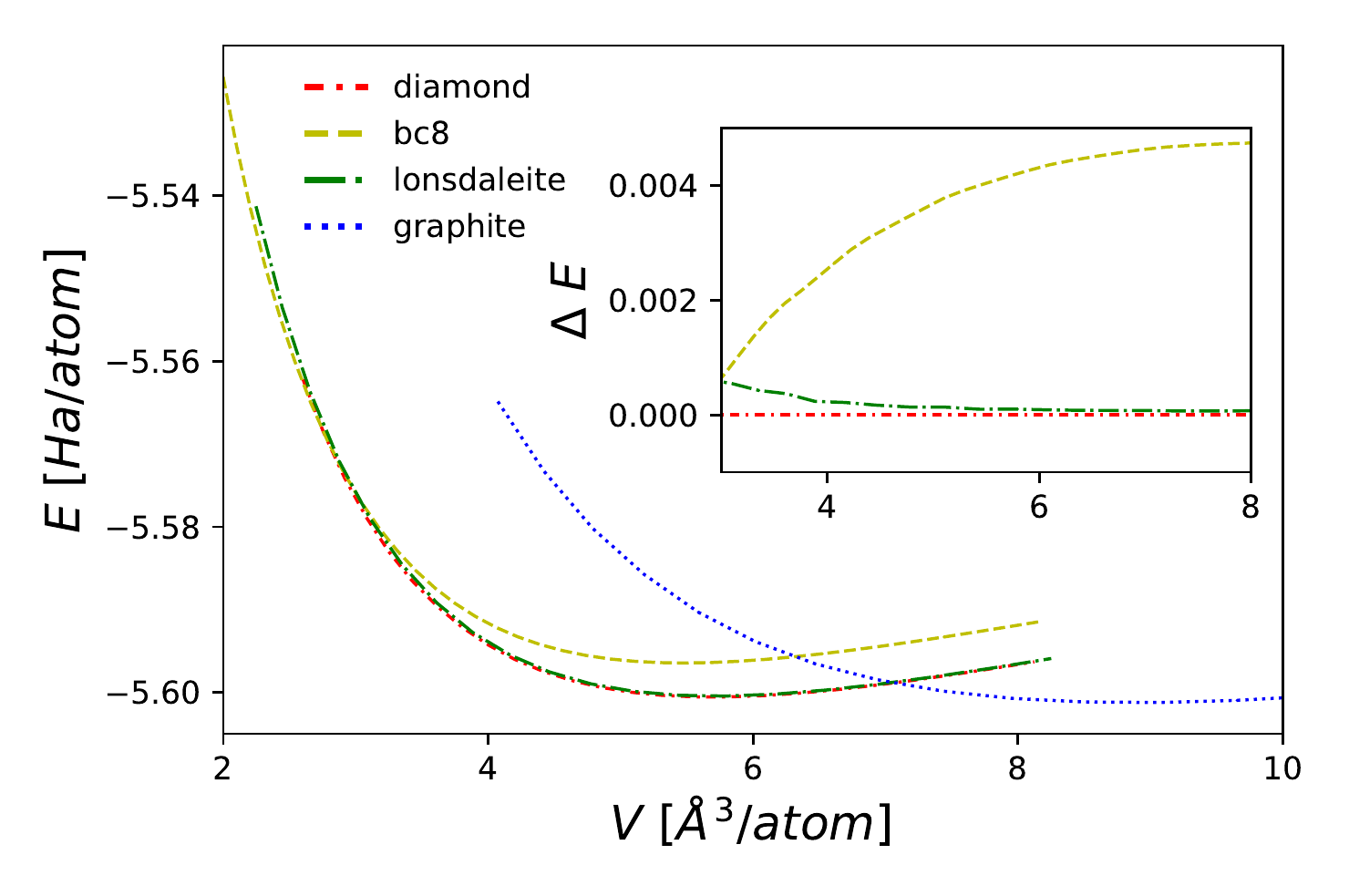}
\caption{ \small{Cold curves for various phases of carbon. The inset panel shows the relative energies with respect to diamond vs volume.} } 
\label{Diamond_eos1}
\end{figure}
 
\begin{figure}[H]
\centering
\includegraphics[width=0.6\columnwidth]{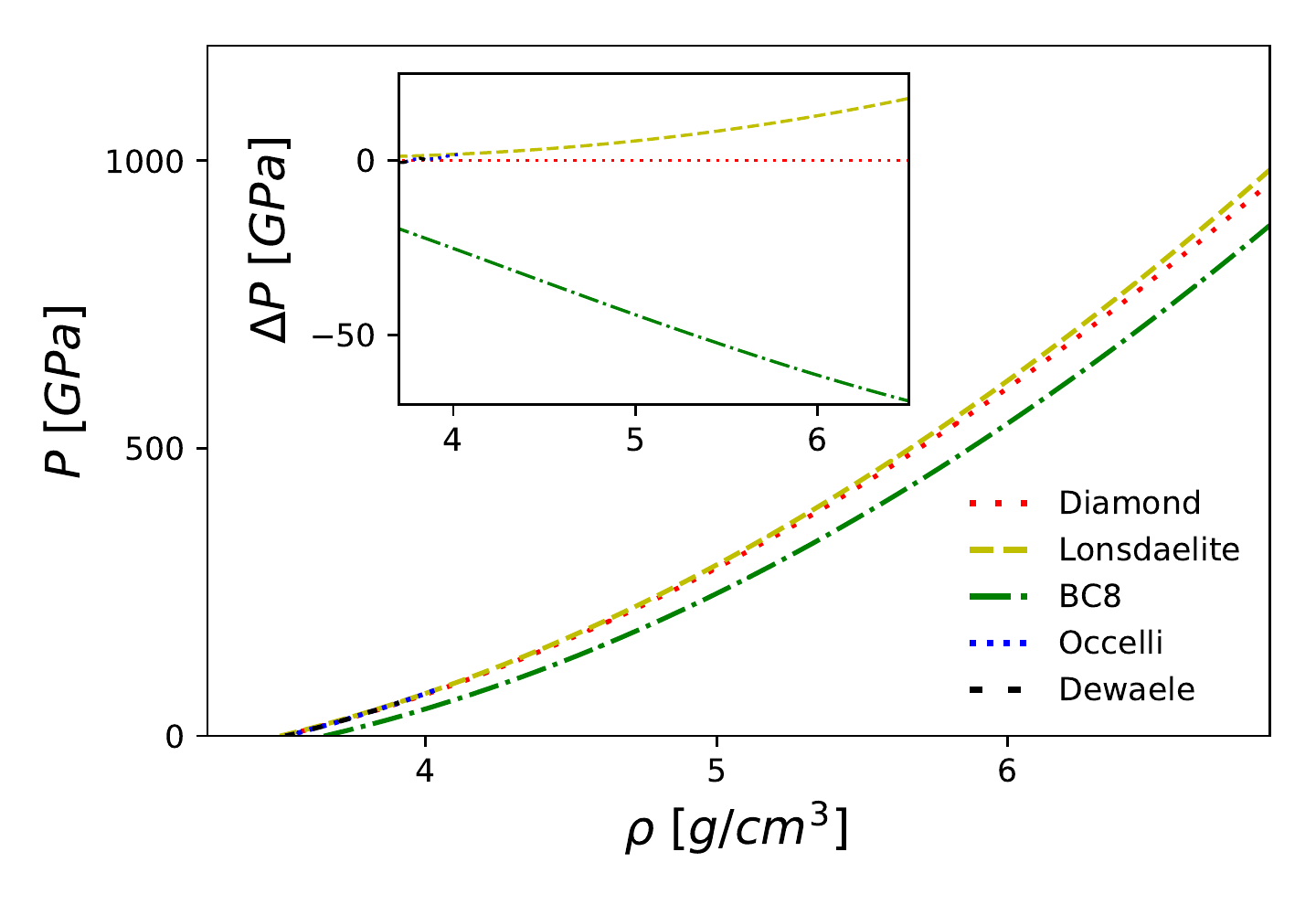} 
\caption{ \small{EOS for various phases in comparison with experiments at 298 K (diamond only). Experimental values by Occelli~\cite{Occelli2003} and Dewaele~\cite{PhysRevB.77.094106} up to 4 $g/cm^{3}$. The inset panel shows the relative difference in pressures with respect to diamond vs density, $\Delta P = P - P_{dia}$. } }  
\label{Diamond_eos2}
\end{figure}

The DFT calculations using elk don't provide the necessary output of pressure. Hence we use the \enquote{eos} utility in elk to first fit the DFT results to the Vinet equation~\cite{0953-8984-1-11-002, PhysRevB.35.1945} to obtain the volume dependent parameters: lattice volume, bulk modulus, pressure derivative and total energy. This provides the corresponding densities and pressures for all our calculations. The cold curves and the EOS for various phases are shown in figures \ref{Diamond_eos1} and \ref{Diamond_eos2} respectively.

The calculated equilibrium parameters are listed in table \ref{t:latticeparameter2} along with the available experimental results.  The band gaps at equilibrium volume using GW$_{0}$ and various xc functionals are summarized in table \ref{t:bandgap}.

\begin{table}[H]
\caption{ \small{ The equilibrium lattice parameters, equilibrium volume per atom, bulk modulus, pressure derivative and static dielectric constants for various phases of carbon. The double row values shown for hexagonal diamond are for the lattice parameters, a$=$b and c. The static dielectric constant is computed using BSE. The values shown in parentheses are experimental data.  
$^{\rm a}$~\cite{Madelung};
$^{\rm b}$~\cite{doi:10.1063/1.1841236};
$^{\rm c}$~\cite{PhysRev.182.891};
$^{\rm d}$~\cite{0953-8984-26-4-045801}.} }
\centering
\begin{tabular}{lccccc}
\hline
\hline
phase  &  a$_{0}$ (\AA) & V$_{0}$ (\AA$^{3}$) & B$_{0}$ (GPa) &  B$_{0}^{'}$ &  $\epsilon_{1}(0)$  \\
\hline 
\hline 
\vspace{0.4cm}
bcc & 4.437 & 5.46  & 433.1 & 3.97 & 8.04 \\
fcc & 3.569 (3.567)$^{\rm a}$ & 5.68  & 430.8 & 3.82 & 5.76 (5.9)$^{\rm c}$ \\
hex. & 2.524 (2.52)$^{\rm b}$ & 5.69  & 430.9 & 3.87  & 6.25 (6.31)$^{\rm b}$ \\
& 4.128 (4.12)$^{\rm b}$ &  &  &   & 5.40 (5.79)$^{\rm d}$ \\
\hline
\hline
\end{tabular}
\label{t:latticeparameter2} 
\end{table}

\begin{table}[H]
\caption{ \small{Band gap (indirect) results in eV. We use GW$_{0}$ with HSE for the fcc phase and PBE for all the phases. $^{\rm a}$~\cite{Cardona}. } }
\centering 
\begin{tabular}{lccccccc}
\hline
\hline
phase & GW & HSE06 & PBE0 & B3LYP & SCAN & PBE & Exp.   \\
\hline
\hline
\vspace{0.2cm}
bcc & 1.94 & 1.89 & 2.64 & 2.24 & 1.11 & 0.90  & -    \\
fcc  & 5.47 & 5.31 & 6.04 & 5.61 & 4.56 & 4.19  & 5.48$^{\rm a}$ \\
hex.  & 4.65 & 4.66 & 5.40 & 4.93 & 3.86 & 3.48   & -  \\
\hline
\hline
\end{tabular}

\label{t:bandgap}
\end{table}

\section{The influence of the xc-kernel in TDDFT}
\label{app_tddft_xc}

\begin{figure}[H]
\centering
\includegraphics[width=0.6\columnwidth]{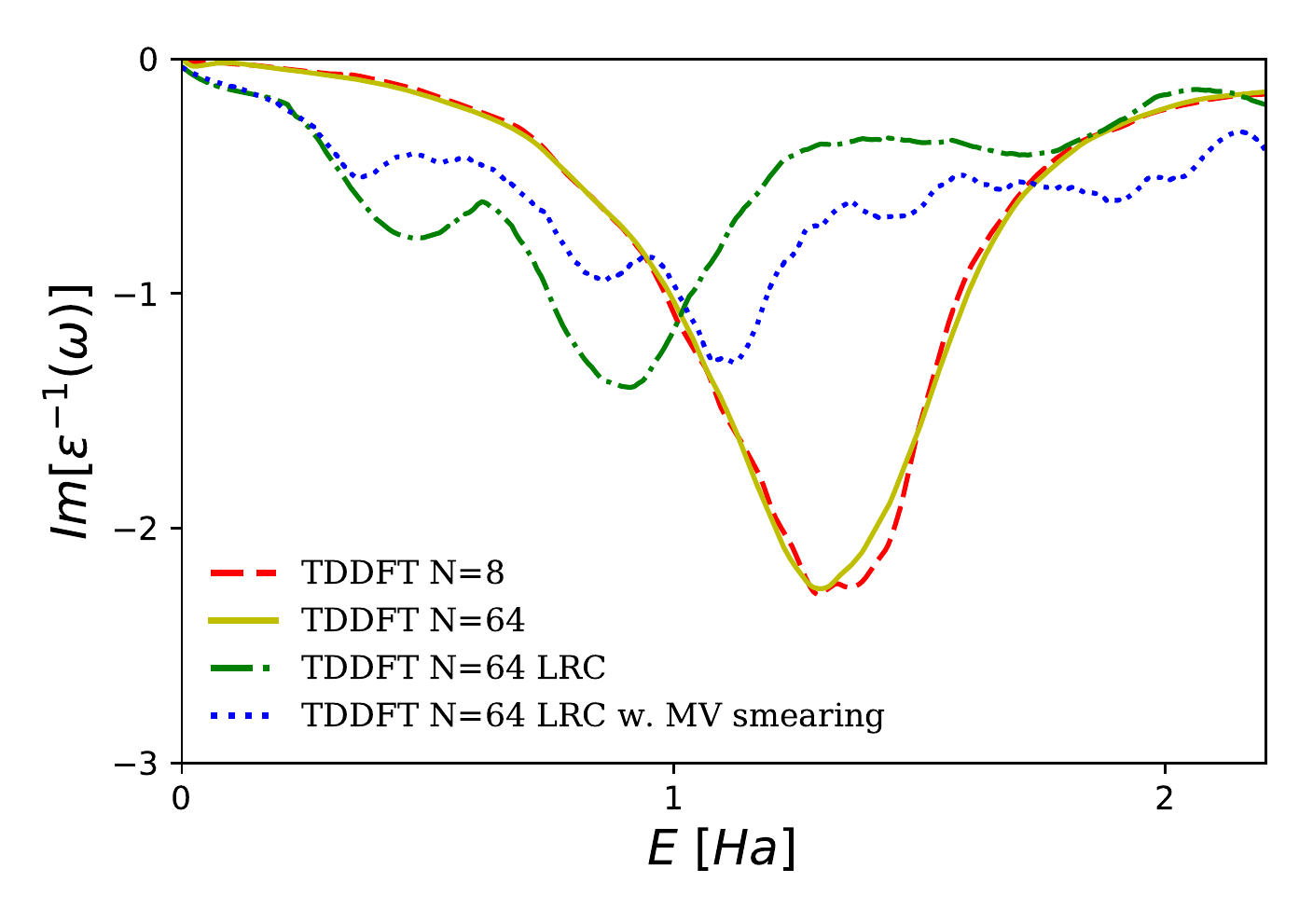} 
\caption{ \small{The change in the imaginary part of the inverse dielectric function of diamond at $p=150$ GPa and $T=6000$ K with system size and different xc kernels in TDDFT. The Bootstrap kernel is used when not labeled otherwise. LRC is the long-range contribution kernel~\cite{PhysRevB.69.155112, PhysRevB.72.125203}, MV is Marzari-Vanderbilt smearing~\cite{PhysRevLett.82.3296}} }
\label{wdm_carbon_tddft_conv} 
\end{figure}

The LRC kernel is computed using Quantum ESPRESSO and yambo using an uniform $4 \times 4 \times 4$  \emph{k}-point mesh with 144 bands and an energy cutoff of 40 Ha with a PBE pseudopotential. A Marzari-Vanderbuilt smearing of 0.019 Ha is also used for the electronic temperature due to the presence of the tiny band gap from the DFT calculations. For the bootstrap calculation, due to the use of the all-electron code elk, the \emph{k}-point mesh is reduced to $3 \times 3 \times 3$. In figure \ref{wdm_carbon_tddft_conv}, we show the convergence of the TDDFT result for warm dense carbon. Contrary to the KG result, the TDDFT result depends strongly on the system size. $N=64$ particles is the most we can afford on the available compute infrastructure. We note the strong influence of the xc kernel used in the TDDFT procedure. The LRC kernel uses free parameters, whereas the Bootstrap kernel avoids these via a self consistency procedure~\cite{PhysRevB.72.125203, PhysRevLett.107.186401}. The cold smearing may be used to speed up convergence as the band gap in warm dense carbon at these conditions is very small. However, this strongly changes the response function.  
 
\section{Lonsdaleite structure}
\label{lons_appendix}

\begin{figure}[H]
\begin{minipage}[c]{1.0\columnwidth} 
\centering
\subfloat[\label{lons_param:a2}]{ \includegraphics[width=0.5\columnwidth]{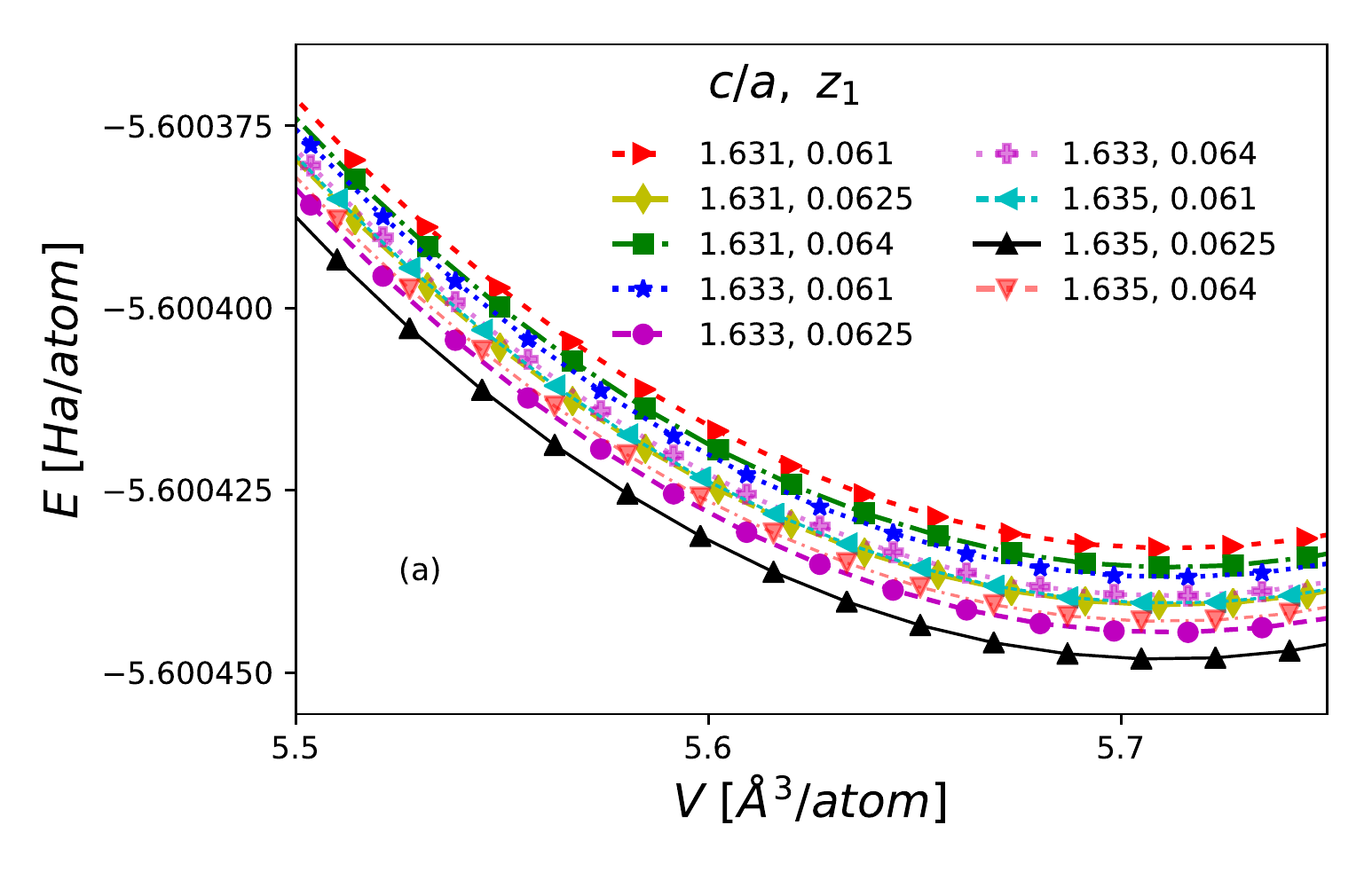} }\par 
\subfloat[\label{lons_param:b2}]{ \includegraphics[width=0.5\columnwidth]{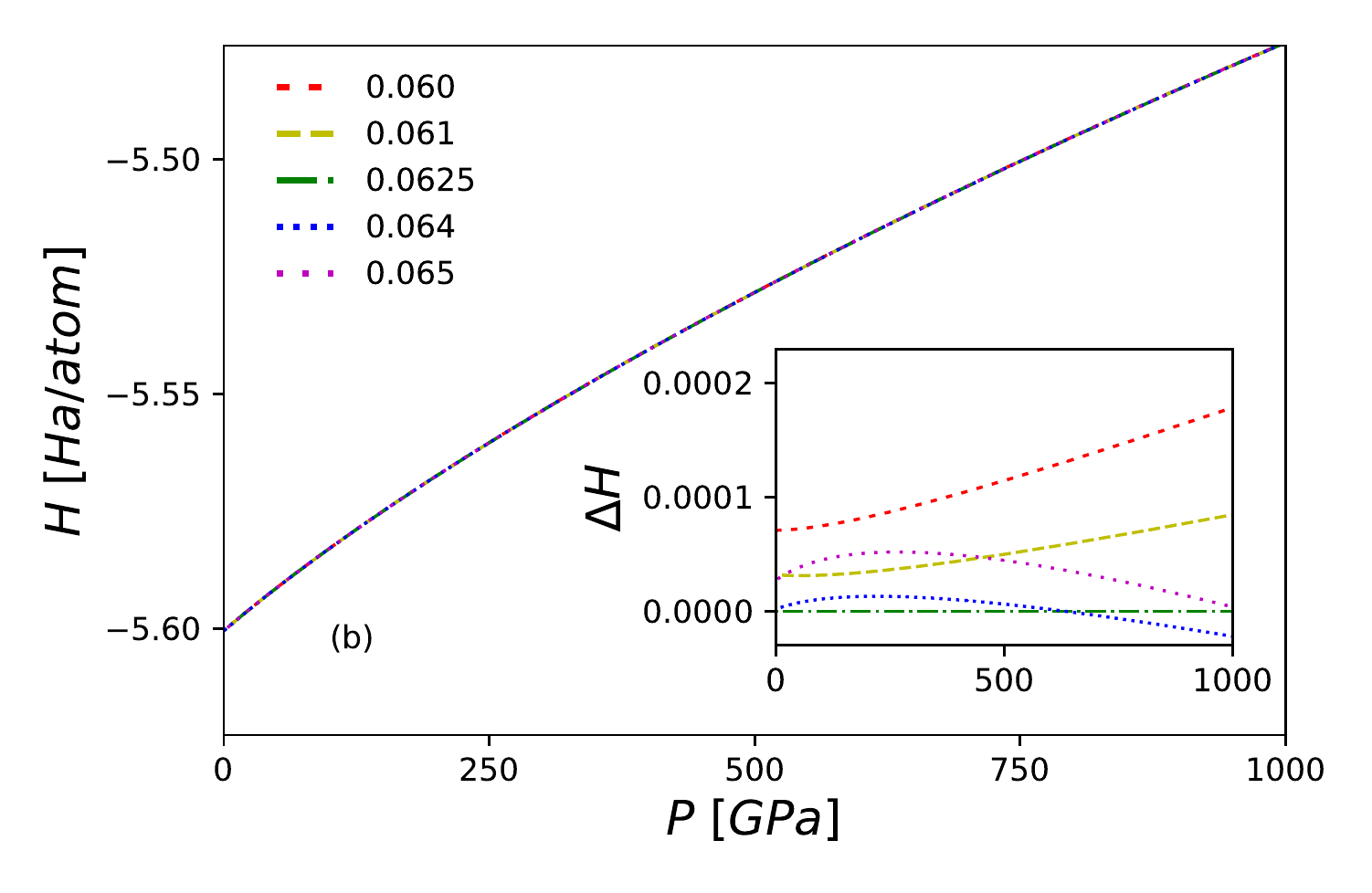} }
\caption{ \small{(a) Variation of the parameters $c/a$ and $z_{1}$ for lonsdaleite. (b) Enthalpy vs pressure with respect to $z_{1}$ for lonsdaleite. The inset plot shows the relative enthalpies with respect to $z_{1}=0.0625$ vs pressure for $c/a=1.635$. } } 
\label{lons_z_parameter2} 
\end{minipage}
\end{figure}

In figure \ref{lons_param:a2}, we show the variation in energy with respect to volume for different values of $z_{1}$ and $c/a$.  The ideal value of $z_{1}=0.0625$ for $c/a=1.635$ is favorable up to $600$ GPa based on the enthalpy changes for the equilibrium lattice parameters in figure \ref{lons_param:b2}.  

\section{BC8 structure}
\label{bc8_appendix}

\begin{figure}[H]
\begin{minipage}[c]{1.0\columnwidth}
\centering
\subfloat[\label{fig:12a2}]{ \includegraphics[width=0.5\columnwidth]{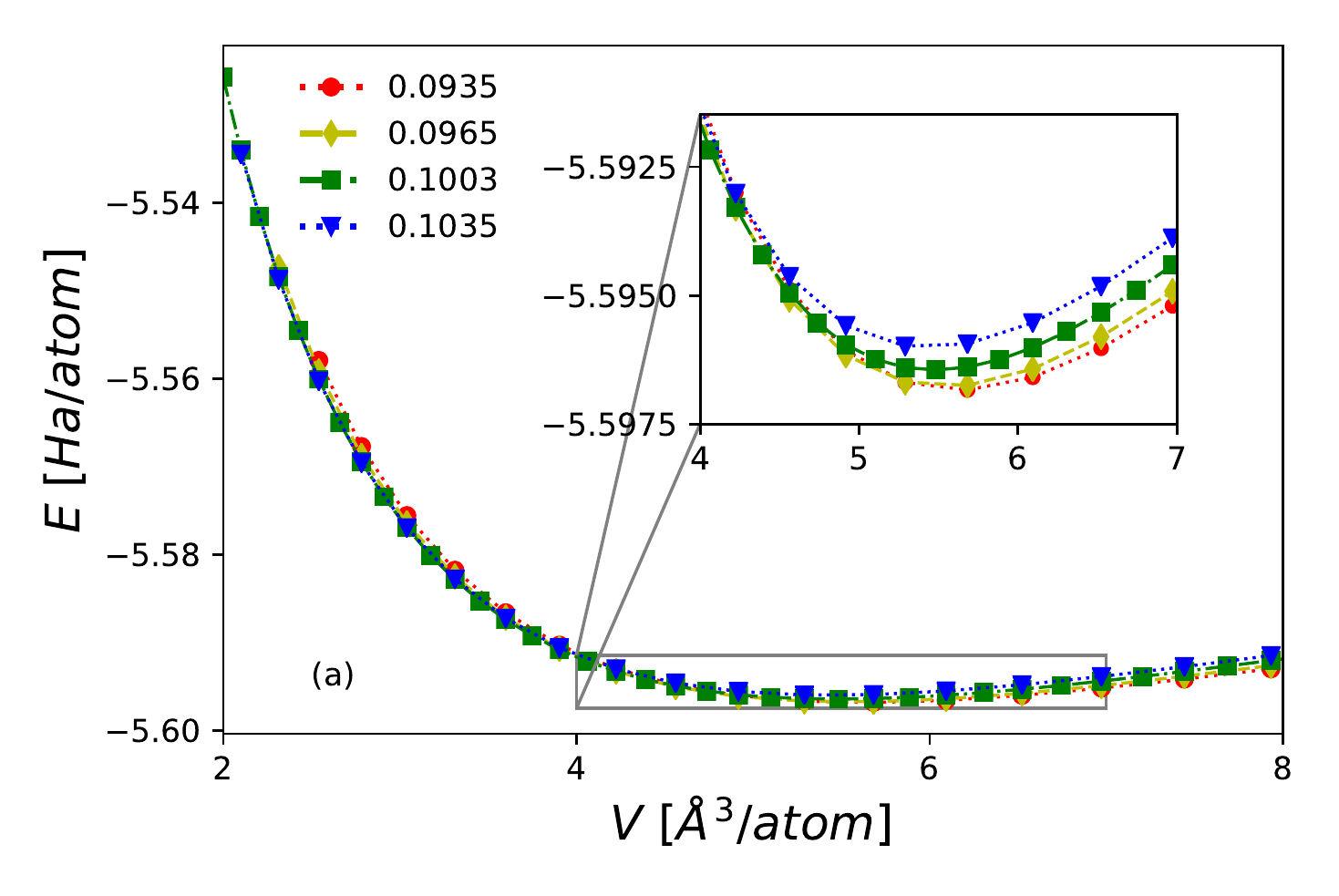} }\par
\subfloat[\label{fig:12b2}]{  \includegraphics[width=0.5\columnwidth]{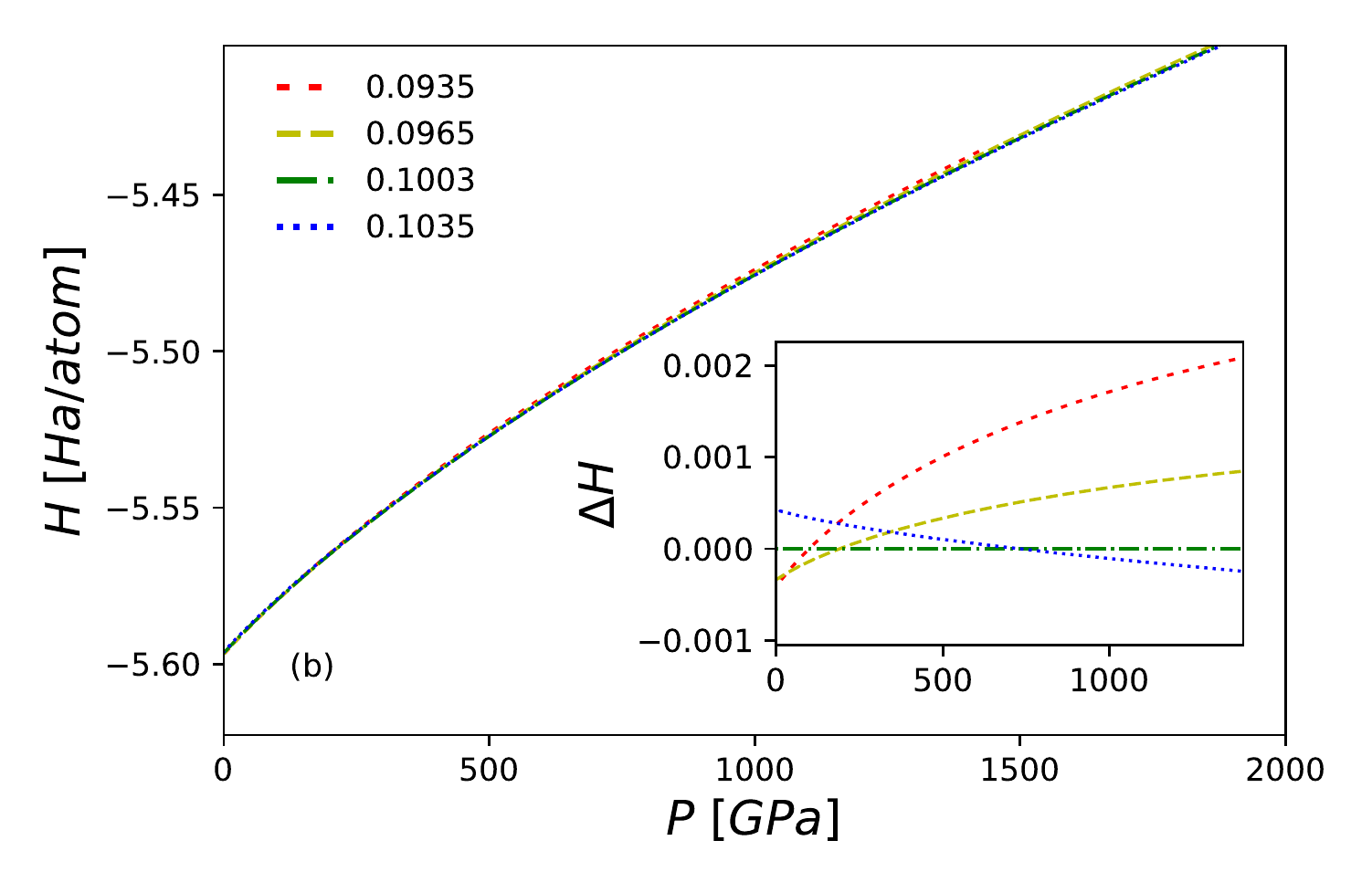} }
\caption{ \small{(a) Variation of the parameter $x_{1}$ for BC8. The inset panel zooms at the region of the minimum. (b) Enthalpy vs pressure with respect to $x_{1}$ for bc8. The inset plot shows the relative enthalpies with respect to $x_{1}=0.1003$ vs pressure. $x_{1}=0.935$ suggested by Clark is ideal for the formation of bc8 phase from diamond and at higher pressures the larger internal parameter is better suited~\cite{ClarkStewart}.} } 
\label{bc8_z_parameter_b22}
\end{minipage}  
\end{figure}

In figure \ref{fig:12a2}, we show the variation in energy with respect to volume for different values of $x_{1}$. The enthalpy changes up to 2500 GPa are shown in figure \ref{fig:12b2} for various $x_{1}$ values.

\end{document}